\documentclass[aps,prb,
amsmath,amssymb,
reprint
]{revtex4-1}

\usepackage{graphicx}
\usepackage{bm}
\usepackage{amsmath}
\usepackage{relsize}
\usepackage{xspace}
\usepackage{booktabs}
\usepackage[version=3]{mhchem}
\usepackage[utf8]{inputenc}
\usepackage[tight]{units}
\usepackage{enumitem}
\usepackage{xcolor}

\usepackage{algorithm}
\usepackage{algpseudocode}

\newcommand{\vc}{v_c}
\newcommand{\LDOS}{{\rm LDOS}}

\newcommand{\epsadj}{\ensuremath\varepsilon^{\dagger}}
\newcommand{\Laplace}{\nabla^2}%
\newcommand{\D}{\mathop{}\!\mathrm{d}}
\newcommand{\cconj}[1]{{#1}^\ast}
\renewcommand{\vec}{\mathbf}
\newcommand{\kk}{\vec{k}}
\newcommand{\GG}{\vec{G}}
\newcommand{\qq}{\vec{q}}
\newcommand{\rr}{\vec{r}}

\begin{document}
\title{
Black-box inhomogeneous preconditioning for self-consistent field iterations in density
functional theory
}
\author{Michael F. Herbst}
\email{michael.herbst@inria.fr}
\affiliation{
  Inria Paris and CERMICS, Ecole des Ponts, 6 \& 8 avenue Blaise Pascal, 77455 Marne-la-Vallee, France.
}

\author{Antoine Levitt}
\email{antoine.levitt@inria.fr}
\affiliation{
  Inria Paris and CERMICS, Ecole des Ponts, 6 \& 8 avenue Blaise Pascal, 77455 Marne-la-Vallee, France.
}

\begin{abstract}
  We propose a new preconditioner based on the local
  density of states for computing the self-consistent
  problem in Kohn-Sham density functional theory. This preconditioner is inexpensive and able to
  cure the long-range charge sloshing known to hamper convergence in
  large, inhomogeneous systems such as clusters and surfaces. It is based on a
  parameter-free and physically motivated approximation to the
  independent-particle susceptibility operator, appropriate for both metals
  and insulators. It can be extended to semiconductors by using the
  macroscopic electronic dielectric constant as a parameter in the model. We
  test our preconditioner successfully on inhomogeneous systems
  containing metals, insulators, semiconductors and vacuum.
\end{abstract}

\maketitle

\section{Introduction}

Since its original formulation in the 1960s~\cite{Hohenberg1964,Kohn1965}
Kohn-Sham density-functional theory~(DFT)
has become one of the most widespread methods
for simulating properties in solid-state physics and chemistry.
Its advantageous balance balance between accuracy and computational cost
as well as the continuous development of
more efficient methods~\cite{kresse1996efficient}
has continuously increased its range of applicability.
Nowadays DFT-based \textit{ab initio}
treatments are standard in many fields~\cite{Norskov2011,Hasnip2014}.
Driven by the availability of
ever-growing amounts of computational power,
several groups have developed high-throughput DFT frameworks where the
properties of a large numbers of systems are computed systematically~\cite{Morgan2005,Jain2011,Setyawan2011}.
In domains such as
catalysis~\cite{Greeley2006,Studt2008,Skulason2012,Johnson2020}
or battery research~\cite{Hautier2011},
where systematic experiments are expensive or time-consuming,
large-scale screening studies represent an interesting alternative
to computationally discover novel compounds
or boil down a large candidate list
to a tractable set for more detailed follow-up investigation.
Compared to the early years where the focus was on performing
a small number of computations on hand-picked systems,
high-throughput studies have much stronger requirements.
In particular any form of manual parameter selection is not practical.
The typical approach to tackle this problem
is to provide a set of heuristics,
which automatically select the computational parameters
based on prior experience~\cite{Curtarolo2012}.
This empirical process is, however, problematic,
because resulting methods can still fall short or break down in cases
with unexpected behavior,
amongst which one often finds exactly the interesting cases
such studies were hoped to excavate in the first place.
A clear objective for improving DFT algorithms
is therefore to employ as small a number of parameters
as possible,
ideally leading to a simulation workflow being composed
only of black-box building blocks that
make use of the appropriate physics
to adapt to the system at hand.

With this idea in mind, this work will consider
the self-consistent field~(SCF) problem as a fundamental step
in all Kohn-Sham DFT simulations.
As it is well-known, simple SCF schemes
suffer from a number of instabilities, which in the linear regime can
be seen to arise from the properties of
the dielectric operator $\varepsilon = 1 - (\vc + K_{\rm XC}) \chi_{0}$.
This operator is built up from three terms: the
independent-particle susceptibility $\chi_{0}$, the Coulomb kernel
$\vc$ and the exchange-correlation kernel $K_{\rm XC}$. Based on an
analysis of these three components, we classify the possible sources of
slow convergence in three types: (a)
systems close to an electronic phase transition, for instance
ferromagnetic, due to the negativity of $K_{\rm XC}$;
(b) localized $d$ or $f$ orbitals, giving large contributions to $\chi_{0}$;
or (c) the long-range ``charge-sloshing''
due to the divergence of the Coulomb operator $\vc$ at large
wavelengths, especially prominent as larger systems are considered.
The latter forms the main focus of this work.
For an early discussion of these aspects, we refer to
the work of \citet{dederichs1983self}.
In applications such as the ones mentioned above,
all these sources of instabilities may occur
and need to be accounted for by a preconditioner
to yield fast and reliably converging SCF methods
for high-throughput calculations.
Moreover, in fields such as catalysis
one is typically concerned with
interfaces, surface effects or surface defects,
i.e.~situations in which simulations
on large and/or inhomogeneous systems are to be performed.

Based on the successful application of the preconditioner
suggested by \citet{Kerker1981} for bulk metallic systems
and (with minor modifications)
for bulk semiconductors~\cite{kresse1996efficient},
several attempts have been made to generalize Kerker's approach to inhomogeneous systems.
For example, the work of \citet{raczkowski2001thomas}
discusses an approach by solving a Thomas-Fermi-von Weizsäcker problem
at every SCF step.
More recently, \citet{zhou2018applicability}
suggested a more generalized parametrization of the Kerker model
for mixed systems along with various strategies to adjust
the respective parameters.
A more general approach,
solving a partial differential equation with variable coefficients, is
the ``elliptic preconditioner'' by~\citet{lin2013elliptic}.
Related is the work by \citet{hasnip2015auxiliary}
where SCF convergence is improved by mapping at each SCF step the Kohn-Sham
problem to a simpler ``auxiliary system'' and solving that.
Yet another alternative approach
is to directly compute an approximate dielectric matrix
based on the Kohn-Sham orbitals%
~\cite{ho1982dielectric,anglade2008preconditioning}
and use this operator for preconditioning. We refer to
\citet{woods2019computing} for a recent and thorough review of these approaches.

While these methods have been successfully applied to different
classes of inhomogeneous systems like metallic slabs or clusters, none
of them has so far proven to be both robust and general enough to
prevent charge-sloshing in all kinds of inhomogeneous systems.
Additionally some of these methods can be either difficult to
implement, too expensive to employ in big systems or require the
selection of many or complicated system-specific parameters, making
automated computations non-trivial. In this work, we present an
inexpensive and easy to implement
preconditioning strategy based on the local density of states. For
systems involving only metals, insulators and vacuum, this
preconditioner is parameter-free and thus completely black-box. Its
range of applicability can be extended to systems involving
semiconductors by introducing an additional parameter, the
dielectric constant --- which for most materials is \textit{a priori}
known. For all tested systems, we find that our preconditioner is able
to cure long-range charge sloshing (cause (c) in the above
classification), leaving only localized states and
electronic phase transitions as sources of slow convergence to be
investigated in future work.

The structure of this manuscript is as follows.
Section~\ref{sec:scf} reviews the mathematical structure
of the SCF problem in Kohn-Sham DFT and provides key results
about rates of convergence,
which are illustrated by example calculations
in Section~\ref{sec:Anderson}.
Section~\ref{sec:chi0} reviews common strategies
for SCF preconditioning in bulk systems
for which Section~\ref{sec:comphomo} provides computational results.
Finally Sections~\ref{sec:prechetero} and~\ref{sec:comphetero}
introduce the LDOS preconditioner in contrast to other
approaches for preconditioning heterogeneous systems
and presents an extensive study illustrating its performance
in large mixed systems.

\section{Self-consistent field iterations}
\label{sec:scf}
Assume an isolated system of $2N$ electrons with periodic boundary
conditions in a (possibly nonlocal) pseudopotential $V_{\rm ext}$ created by
nuclei and core electrons. The Kohn-Sham equations
for $N$ electron pairs at finite temperature $T$ are (in atomic units)
\begin{equation}
\begin{aligned}
  &\left( -\frac 1 2 \Laplace + V_{\rm ext} + V_{\rm HXC}(\rho) \right) \psi_{i} = \varepsilon_{i} \psi_{i}\\
  &\rho = \sum_{i = 1}^{\infty} f\left(\frac{\varepsilon_{i} - \varepsilon_{F}}{T}\right) |\psi_{i}|^{2}\\
  &V_{\rm HXC}(\rho) = \vc \rho + V_{\rm XC}(\rho)\\
  &\int \rho(\rr) \D\rr = 2 N
\end{aligned}
\end{equation}
where $f(x) = 2/(1+e^{x})$ is twice the Fermi-Dirac function, $V_{\rm XC}(\rho)$ is
the exchange-correlation potential, and $\vc \rho$ is
the Coulomb potential associated to the charge density $\rho$ in a
uniform neutralizing background:
\begin{align}
  -\Laplace (\vc \rho) = 4\pi \rho.
\end{align}
We define the potential-to-density mapping $F$ by
\begin{align}
  F(V) &= \sum_{i=1}^{\infty} f\left(\frac{\varepsilon_{i} - \varepsilon_{F}}{T}\right) |\psi_{i}|^{2}
\end{align}
where
$(\varepsilon_{i},\psi_{i})$ are the
eigenpairs of $-\frac 1 2 \Laplace + V_{\rm ext} + V$ and
the Fermi level $\varepsilon_F$ is uniquely defined by the condition
\begin{equation}\int \Big(F(V)\Big)(\rr) \D\rr = 2N.\end{equation}
The Kohn-Sham
equations can then be written as the fixed-point equation
\begin{align}
  \label{eq:SCF_eq}
  F(V_{\rm HXC}(\rho)) = \rho.
\end{align}
The simplest self-consistent iteration is
\begin{align}
  \rho_{n+1} = F(V_{\rm HXC}(\rho_{n})).
\end{align}
Except on very simple systems, this iteration usually does not converge.
A more useful scheme is the \textit{damped} iteration
\begin{align}
  \label{eq:simple_SCF}
  \rho_{n+1} = \rho_{n} + \alpha \left[F\big(V_{\rm HXC}(\rho_{n})\big) - \rho_{n}\right],
\end{align}
where $\alpha > 0$ is a damping parameter. Note that we consider in
this paper density mixing algorithms, but all our results are easily
transposable to potential mixing.

It is very instructive to consider the convergence behavior of this
iteration near a fixed-point $\rho_{*}$. As is well-known, the
iteration converges locally if and only if all the eigenvalues of the Jacobian matrix
\begin{align}
  J_{\alpha} = 1 - \alpha (1 - \chi_{0} K)
\end{align}
have magnitude less than one. In this case, the convergence rate is equal to
the eigenvalue of largest magnitude of $J_{\alpha}$. In this equation,
\begin{align}
  \chi_{0} = F'\big(V_{\rm HXC}(\rho_{*})\big)
\end{align}
is the independent-particle susceptibility, and
\begin{align}
  K = V_{\rm HXC}'(\rho_{*}) = \vc + K_{\rm XC}(\rho_{*})
\end{align}
is the derivative of the Hartree-exchange-correlation potential (the
Hartree-exchange-correlation kernel).

Note that $J_{\alpha} = 1 - \alpha \epsadj$, where
\begin{align}
  \epsadj = 1 - \chi_{0} K
\end{align}
is the adjoint of the dielectric matrix
$\varepsilon = 1 - K \chi_{0}$. As will be shown in Section \ref{sec:chi0}, the susceptibility
$\chi_{0}$ is a non-positive self-adjoint operator, and $K$ is self-adjoint. It follows that
\begin{equation}
  \chi_{0} K = - (-\chi_{0})^{1/2}(-\chi_{0})^{1/2} K
	\label{eqn:similar}
\end{equation}
has the same spectrum (except possibly for $0$) than the symmetrized
self-adjoint operator $(-\chi_{0})^{1/2} K (-\chi_{0})^{1/2}$, and
therefore that $\epsadj$ has real spectrum.

It is often a good approximation to neglect the term
$K_{\rm XC} = V_{\rm XC}'(\rho_{*})$ in $K$, yielding the random phase
approximation (RPA)
\begin{align}
  K_{\rm RPA} = \vc.
\end{align}
Since $\vc$ has non-negative spectrum, $\epsadj = 1 - \chi_{0} \vc$
has positive spectrum. If we do not assume the RPA approximation,
$K$ does not have a guaranteed sign due to the non-convexity
introduced in the model by the exchange-correlation term.
The eigenvalues of $\chi_{0} K$ can thus be in principle arbitrary.
It can, however, be
shown that, at an energy local minimizer, $\epsadj$ has non-negative
spectrum \cite{dederichs1983self,gonze1996towards,cances2020convergence}.

It follows that, near a non-degenerate energy minimizer, by selecting
$\alpha$ small enough the spectrum of
$J_{\alpha} = 1 - \alpha \epsadj$ will be in $(-1,1)$, and therefore
the damping iteration \eqref{eq:simple_SCF} is locally convergent (see
\citet[Theorem 3.6]{cances2020convergence}). However, this convergence
may be unacceptably slow in practice. Letting $\lambda_{\min} > 0$ and
$\lambda_{\max}$ be the lowest and largest eigenvalues of $\epsadj$,
the convergence rate is
$R = \max(|1 - \alpha \lambda_{\min}|, |1- \alpha \lambda_{\max}|)$.
Assuming that $\alpha$ is selected optimally to minimize this
convergence rate, we obtain
\begin{align}
  \label{eq:optimal_step}
  \alpha = \frac 2 {\lambda_{\min} + \lambda_{\max}}
\end{align}
and
\begin{equation}
	R = \frac{\kappa-1}{\kappa+1},
	\label{eqn:RateSimple}
      \end{equation}
      where
      \begin{align}
        \label{eq:condition_number}
  \kappa = \frac{\lambda_{\max}}{\lambda_{\min}}
\end{align}
is the (metric-independent) ``spectral condition number''. Note that
this number is distinct from the usual (metric-dependent) condition
number of $\epsadj$, equal to the ratio of large to lowest
\textit{singular} values, because this operator is not self-adjoint.
The spectral condition number $\kappa$ can however be identified with
the condition number of $\epsadj$ under the metric induced by
$-\chi_0$, for which $\epsadj$ is self-adjoint. With an abuse of
language, we will use the term ``condition number'' to refer to the
``spectral condition number'' \eqref{eq:condition_number} throughout
this work.

For $\kappa$ large enough, $R \approx 1 - 2/\kappa$ and therefore the
number of iterations $\frac{\tt tol}{\log R}$ needed to achieve a
fixed tolerance ${\tt tol}$ grows proportionally to $\kappa$. Slowly
converging SCF iterations are therefore linked to large values of
$\kappa$, originating from small and large eigenvalues of
$\epsadj = 1 - \chi_{0} (\vc + K_{\rm XC})$. In practice, this can be
attributed to three different classes of instabilities, linked to
large modes of $K_{\rm XC}, \chi_{0}$ and $\vc$ respectively:
\begin{enumerate}[label=(\alph*)]
\item The first case is when $\epsadj$ has small eigenvalues. This can
  only happen in systems with strong exchange-correlation effects,
  because under the RPA all eigenvalues of $\epsadj$ are greater than
  $1$. This phenomenon is usually (but not necessarily) associated
  with symmetry breaking. To see this, consider ferromagnetic systems,
  where $\epsadj$ has negative eigenvalues at
  the spin-unpolarized state (a solution of the Kohn-Sham equations
  which is not an energy minimum). Paramagnetic systems close to a ferromagnetic
  transition will display spin-unpolarized energy minima with small
  but positive eigenvalues for $\epsadj$. The modes associated
  with the slow convergence are spin-polarization modes.
\item A second source of instabilities are large eigenvalues of
  $\epsadj$ brought about by large modes of $\chi_{0}$. In
  solids, these modes can for instance be due to localized states,
  such as $d$ or $f$ electrons, close to the Fermi level. In this case
  the associated modes are localized around atoms.
\item A final source are large eigenvalues of $\epsadj$ originating from the
  long-range divergence of the Coulomb interaction (large modes of
  $\vc$). Indeed, if $\widehat \rho(\qq)$ are the Fourier coefficients of $\rho$, then
\begin{align}
  \widehat{(\vc \rho)}(\qq) = \frac{4\pi \widehat \rho(\qq)}{|\qq|^{2}}
\end{align}
which diverges for small $\vec{q}$ (long wavelengths). Depending on whether
this effect is compensated or not by $\chi_{0}$, this can manifest as
large eigenvalues of $\epsadj$. As we will see, no compensation takes
place for metals in large supercells. In this case, the condition
number increases with the square of the system size length. Associated
modes are long-wavelength transfer of charge from one end of the
system to another (``charge sloshing'').
\end{enumerate}
Among these three sources of instabilities, the third one is possibly
the most serious, because there the convergence degrades with system
size.

In this paper, we focus on curing this source of slow
convergence using preconditioning. Preconditioning consists in replacing \eqref{eq:simple_SCF} by
\begin{equation}
    \rho_{n+1} = \rho_{n} + \alpha P^{-1} \Big(F(V_{\rm HXC}(\rho_{n})) - \rho_{n}\Big)
    \label{eqn:Preconditioning}
  \end{equation}
where $P$ is a preconditioner. The Jacobian of this method at
convergence ($\rho = \rho_{*}$) is
\begin{equation}
  J_{\alpha P} = 1 - \alpha P^{-1} \epsadj.
\end{equation}
It follows that the optimal preconditioner $P$ is simply the adjoint
dielectric operator $\epsadj$. Indeed, in this case, when
$\alpha = 1$, the iteration \eqref{eqn:Preconditioning} turns into the standard
Newton method: $J_{P} = 0$, and the method converges in one step in
the linear approximation. In practice however, inverting this
explicitly (or computing its action on a vector) is too costly.
Therefore it is desirable to find a preconditioner $P$ that can be
inverted efficiently and such that \mbox{$P^{-1} (1-K \chi_{0})$} has a
condition number close to 1.

\section{Anderson acceleration}
\label{sec:Anderson}
Before turning our attention to the problem of finding good
preconditioners, we address the convergence behavior of acceleration
techniques, which are commonly used to speed up SCF convergence,
and which we will use in our tests.

Acceleration techniques treat the fixed-point equation
\eqref{eq:simple_SCF} as a black box iteration, and use the history
$\rho_{1}, \dots, \rho_{n}$ in order to extrapolate a better
$\rho_{n+1}$ than the one obtained by the simple (preconditioned)
damping method \eqref{eqn:Preconditioning}. There are many
variants of these, the most used being the method variously known as
Pulay/DIIS/Anderson mixing/acceleration, which we will refer to as
Anderson acceleration. We refer to \citet{woods2019computing} for an extensive review in the context
of Kohn-Sham DFT, and to \citet{chupin2020convergence} and
references within for a mathematical analysis of these acceleration
techniques. In particular, in the linear regime and with infinite
history, Anderson acceleration is known to be equivalent to the well-known GMRES
method to solve linear equations. Assuming non-linear effects to be
negligible, we can expect that Anderson acceleration inherits the
improved convergence properties of Krylov methods \cite{Saad2003}. In
particular, the convergence rate of the GMRES method is
\begin{equation}
  R = \frac{\sqrt \kappa - 1}{\sqrt \kappa + 1},
  \label{eqn:RateAnderson}
\end{equation}
and therefore the number of iterations grows only as the square
root of $\kappa$. Furthermore, an important property of Krylov methods
is that they converge to the exact solution in $N$ steps if the matrix
has $N$ distinct eigenvalues \cite{Saad2003}, and therefore we can
expect Anderson acceleration to be very efficient in the case where
eigenvalues are clustered.

In practice, Anderson methods are truncated to a finite history size,
and the impact of the nonlinearity on the efficiency of the
extrapolation is complex. Unlike with the simple
damping method, the convergence behavior of Anderson acceleration can
not be understood solely from the condition number $\kappa$, and the
number of iterations does not always follow the law predicted by the
rate \eqref{eqn:RateAnderson}.

\begin{figure}
	\centering
	\includegraphics[width=0.9\columnwidth]{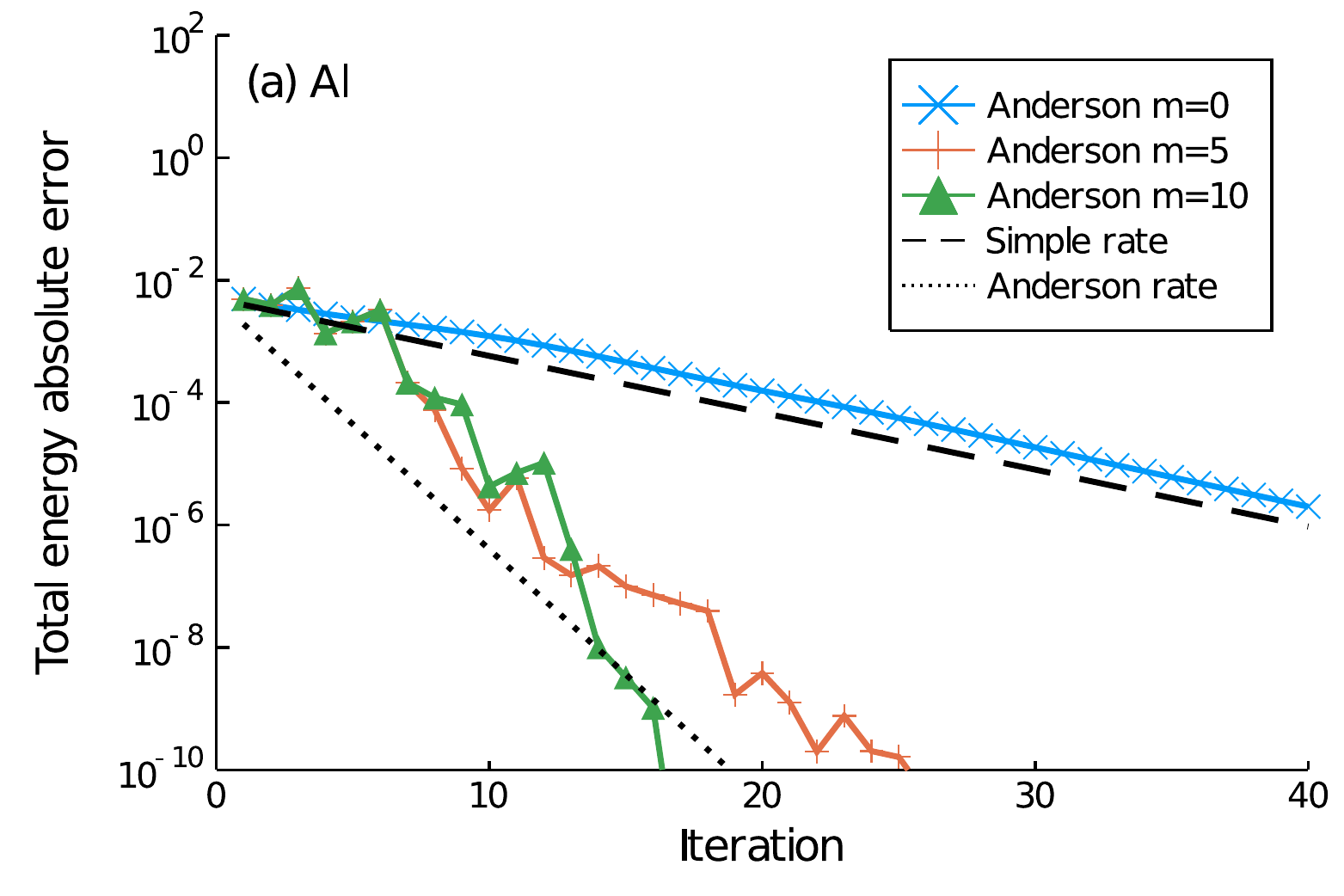}
	\includegraphics[width=0.9\columnwidth]{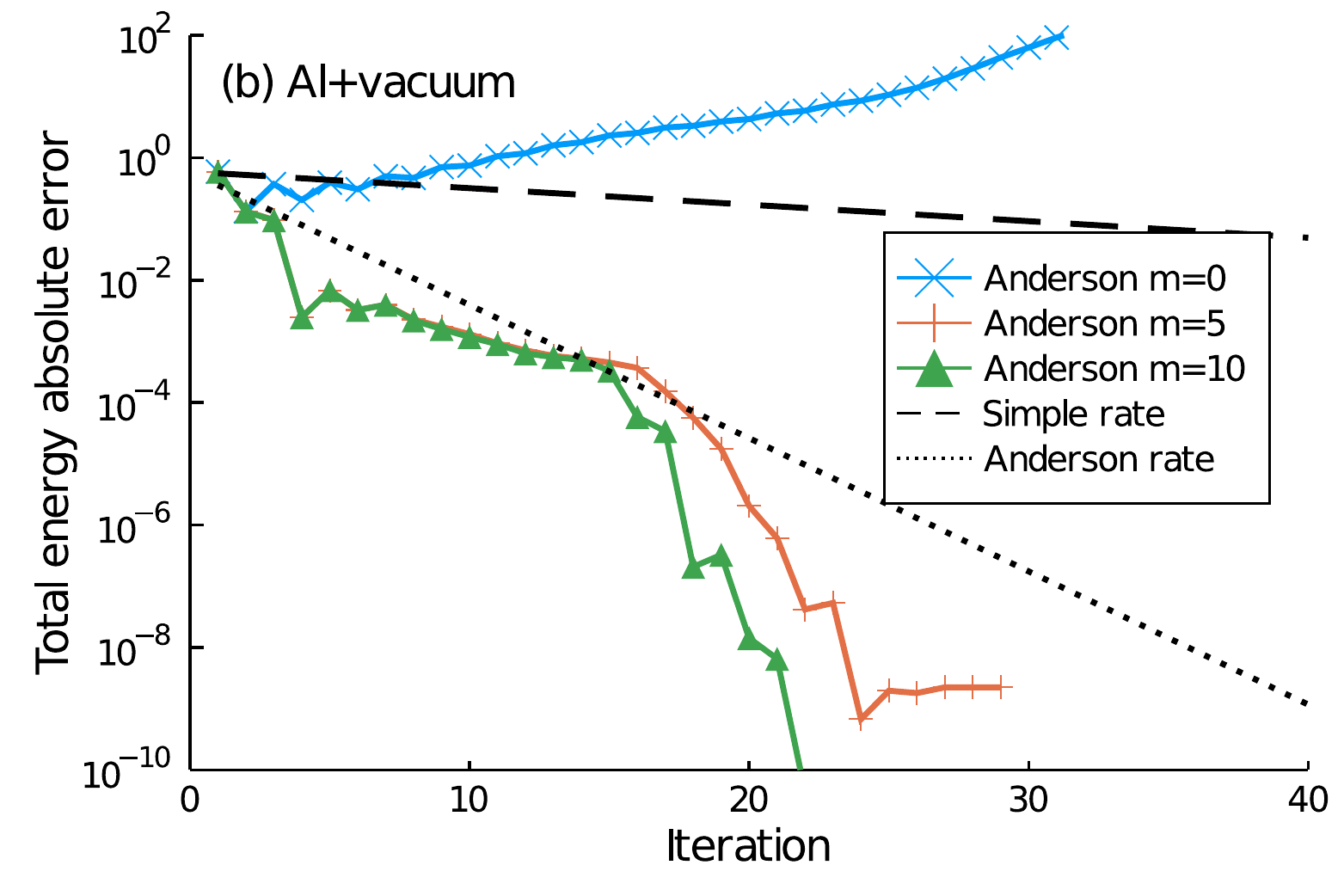}
	\includegraphics[width=0.9\columnwidth]{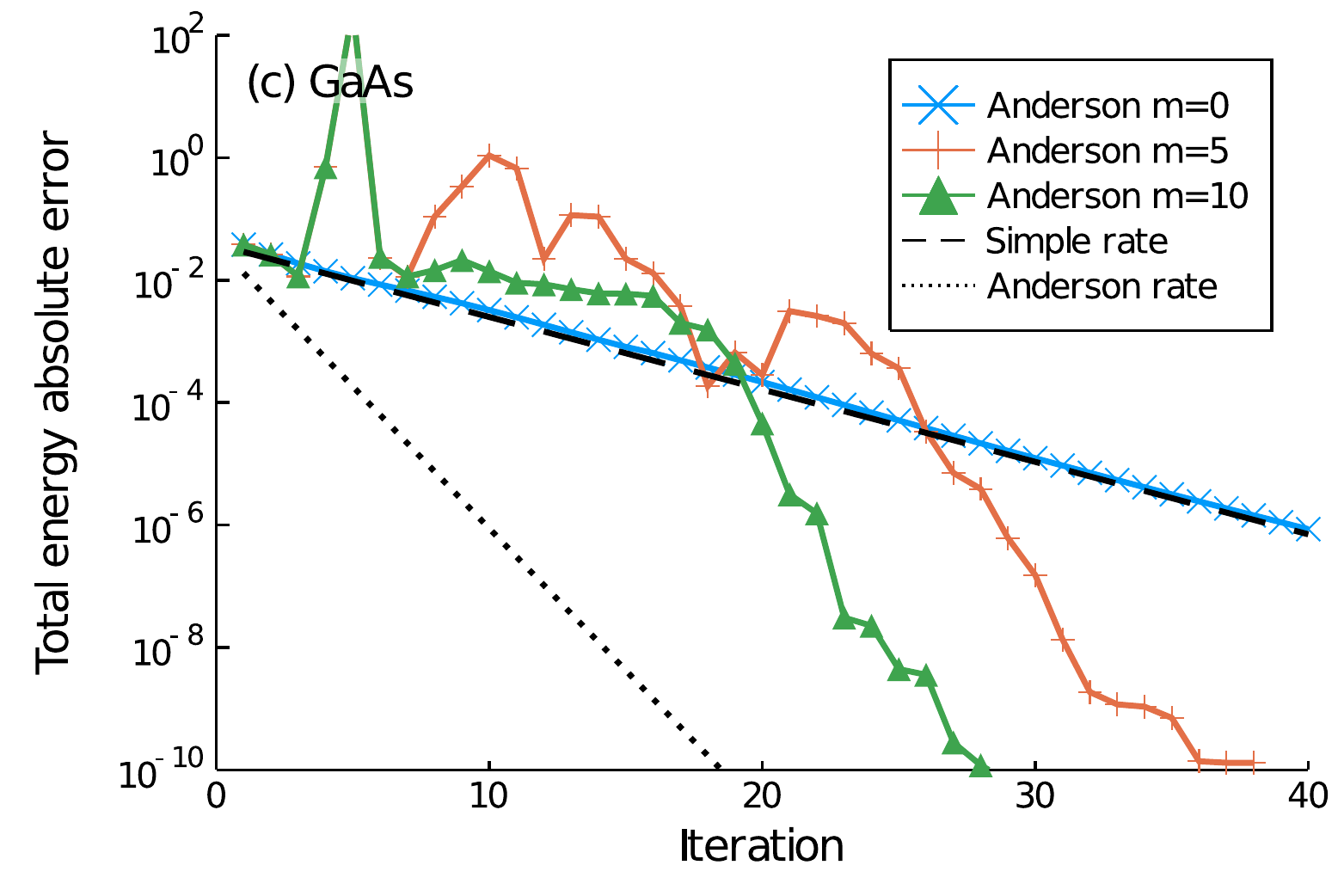}
	\caption{
        Convergence of self-consistent field iterations
        on three selected systems using different history lengths $m$
        for Anderson acceleration ($m=0$ corresponds to simple mixing).
        From top to bottom the considered systems are
        (a) 12 layers of aluminium using a Kerker preconditioner;
        (b) 10 layers of aluminium with an equally wide portion of vacuum;
        and (c) 20 layers of gallium arsenide,
        in both latter cases using no preconditioner.
        }
	\label{fig:anderson}
\end{figure}

In Figure \ref{fig:anderson} we exhibit the error in energy for three
representative cases, selected to illustrate the possible behaviors.
We compute the extreme eigenvalues of $P^{-1} \epsadj$ at convergence,
and use them to determine the optimal step using
\eqref{eq:optimal_step}. For comparison, the plot
also shows the convergence estimated from
the rates \eqref{eqn:RateSimple} for damped iterations and
\eqref{eqn:RateAnderson} for Anderson acceleration.
For details on the computational procedures,
we refer to Sections~\ref{sec:comphomo} and~\ref{sec:comphetero}.

In cases (a) and (c)
the simple mixing method ($m=0$) converges with a rate consistent with
the theoretical prediction. In case (b), using
the optimal step computed at the solution proves too optimistic, and
the iteration is divergent.

The accelerated methods track approximately the convergence rate
\eqref{eqn:RateAnderson} in case (a). Case (b) displays a
characteristic ``plateau'' behavior: after an initial stage of rapid
convergence, the error stagnates, then very rapid convergence is
obtained once the accelerated scheme has ``learned'' the dominant
clusters in the eigenvalues of $P^{-1} \epsadj$. The plateau is
particularly pronounced in this case because $P^{-1} \epsadj$ has
a very clustered spectrum:
the largest eigenvalue is about 50, but all
eigenvalues except five are between 0.9
and 2.0. In case (c) the
error shoots up abruptly around iterations 4 and 5. This is because the
linear model assumed by Anderson acceleration does not hold in the
initial steps. In more complex cases, this can lead to complete
divergence, but here the iteration recovers and then displays the plateau
behavior seen in case (b).
Notice that for $m=5$, where the history size is too small to capture all
dominant clusters, the plateau has more the shape of a chain of half-arcs.

Overall one can conclude that while for non-accelerated fixed-point
iterations the condition number and the resulting theoretical rates
are very closely related to the observed number of iterations, the
situation is less clear for the Anderson-accelerated procedure. In the
following we will still employ Anderson acceleration for all our
calculations for two reasons. Firstly because this represents the
typical situation encountered in practice and secondly because
otherwise, especially for cases with bad preconditioning, convergence
can become extremely slow. To simplify our analysis we will also
largely assume that the condition number of the preconditioned adjoint
of the dielectric matrix $P^{-1}\varepsilon^\dagger$ is a good
indicator for the expected number of iterations in the
Anderson-accelerated SCF, even though this discussion shows
that the correlation is not as clear as
one would expect from the simple estimate \eqref{eqn:RateAnderson}.

Note that, although the Anderson method is sometimes divergent,
convergence can always be achieved (albeit very slowly) with the
simple damping method by selecting a sufficiently small damping
parameter. This means that it should be possible to stabilize the
Anderson method by selecting stepsizes and extrapolation parameters
more conservatively. This is a worthwhile research direction, but falls
outside the scope of this paper.

\section{Response operators and preconditioning}
\label{sec:chi0}
\subsection{The independent-particle susceptibility $\chi_{0}$}
\newcommand{\Ncell}{\mathcal N}
\label{sec:chi0_deriv}

To examine the impact of the Coulomb divergence on the eigenvalues of the
dielectric operator, we study the properties of the
independent-particle susceptibility operator $\chi_{0}$ in periodic systems. Consider a
perfect crystal with lattice $\mathcal R$, reciprocal lattice
$\mathcal R^{*}$, Brillouin zone $\mathcal B$, unit cell $\Omega$ and
$N_{\rm el}$ electron pairs per unit cell. We choose a domain
$\Omega_{\Ncell}$, with periodic boundary conditions, composed
of $\Ncell = L_{1} \times L_{2} \times L_{3}$ copies of the
unit cell of the crystal, and containing $ \Ncell N_{\rm el}$
electron pairs. Then, as is standard, the eigenvectors of the periodic
Hamiltonian $-\frac 1 2 \Laplace + V_{\rm ext} + V_{\rm HXC}(\rho_{*})$
can
be chosen as the Bloch waves
\begin{align}
    \psi_{n\kk}(\rr) = \frac 1 {\sqrt{\Ncell}} e^{i\kk \cdot \rr}
    u_{n\kk}(\rr),
\end{align}
where $u_{n\kk}$ is a $\mathcal R$-periodic function
normalized as
\begin{equation}
\begin{aligned}
    \langle u_{n\kk}|u_{m\kk} \rangle_{\Omega_{\Ncell}}
    &= \Ncell \langle u_{n\kk}|u_{m\kk} \rangle_{\Omega} \\
    &= \Ncell \int_{\Omega} \cconj{u}_{n\kk}(\rr) \, u_{m\kk}(\rr) \D\rr \\
  &= \Ncell \delta_{nm}.
\end{aligned}
\end{equation}
The index $\kk$ runs over the discrete set
$\mathcal B_{\Ncell} \subset \mathcal B$, which contains the $\Ncell$
wavevectors in the Brillouin zone that are compatible with the
periodic boundary conditions on $\Omega_{\Ncell}$.

We now derive an explicit sum-over-states expression for $\chi_{0} = F'(V_{\rm HXC}(\rho_{*}))$. We consider an infinitesimal
perturbation $\delta V$, and compute the response
$\delta\rho = \chi_{0} \delta V$. We get
\begin{align}
  \delta\rho &= \sum_{\kk \in \mathcal B_{\Ncell}}\sum_{n=1}^{\infty} \left( \delta f_{n\kk} \, |\psi_{n\kk}|^{2}
              + f_{n\kk}( \, \cconj{\delta \psi}_{n\kk} \psi_{n\kk}
              +  \, \cconj{\psi}_{n\kk} \delta \psi_{n\kk} )\right)
\end{align}
with $f_{n\kk} = f\left(\frac{\varepsilon_{n\kk} - \varepsilon_{F}}{T}\right)$.
From first-order perturbation theory,
\begin{align}
  \delta f_{n\kk} &=  (\langle\psi_{n\kk}|\delta V \psi_{n\kk} \rangle_{\Omega_{\Ncell}} - \delta \varepsilon_{F}) \ f'_{n\kk} \\
  \delta \psi_{n\kk} &= \sum_{\kk' \in \mathcal B_{\Ncell}} \sideset{}{'}\sum_{m=1}^{\infty} \frac{\langle  \psi_{m\kk'}| \delta V \psi_{n\kk} \rangle_{\Omega_{\Ncell}}}{\varepsilon_{n\kk} - \varepsilon_{m\kk'}} \psi_{m\kk'}
\end{align}
with
\begin{align}
  f'_{n\kk} = \frac 1 T f'\left(\frac{\varepsilon_{n\kk} - \varepsilon_{F}}{T}\right)
\end{align}
and where $\Sigma'$ indicates that the term $m=n$, \mbox{$\kk'=\kk$} is
omitted in the summation.
Note that strictly speaking, these intermediary results are only valid for
non-degenerate eigenvalues (otherwise $\delta \psi_{n\kk}$ is not
well-defined), but the end result is also valid for degenerate
eigenvalues, see \citet{levitt2020screening} for an alternative
derivation using contour integrals. We can determine
$\delta \varepsilon_{F}$ by the condition that the total number of
electrons is unchanged:
$\sum_{\kk \in \mathcal B_{\Ncell}} \sum_{n=1}^{\infty} \delta f_{n\kk} = 0$ and therefore
\begin{align}
  \delta \varepsilon_{F} = \frac{\langle D_{\rm loc}| \delta V \rangle_{\Omega_{\Ncell}}}{\Ncell D}
\end{align}
where we have introduced
\begin{align}
  D &= -\frac 1 {\Ncell |\Omega|} \sum_{\kk \in \mathcal B_{\Ncell}}\sum_{n=1}^{\infty} f'_{n\kk}\\
  D_{\rm loc}(\rr) &= -\frac 1 {\Ncell |\Omega|}\sum_{\kk \in \mathcal B_{\Ncell}}\sum_{n=1}^{\infty} f'_{n\kk} |u_{n\kk}(\rr)|^{2}
\end{align}
the total density of states per unit volume at the Fermi level, and the
local density of states.

We can now obtain the integral kernel of $\chi_{0}$ both in real space and in reciprocal
space. For the former, we simply identify using the formula
$\delta\rho(\rr) = \int_{\Omega_{\Ncell}} \chi_{0}(\rr,\rr') \delta V(\rr') \D\rr'$.
For the latter, note that $\chi_{0}$ is invariant by the
lattice translations $\mathcal R$, and can therefore be described by
its Bloch matrix $\chi_{0}(\qq,\GG,\GG')$, where
$\qq \in \mathcal B_{\Ncell}, \GG, \GG' \in \mathcal R^{*}$,
describing the density response at wavelength $\qq+\GG$ to a
total potential perturbation at wavelength $\qq+\GG'$. Using
the convention
$\frac{f_{n\kk}-f_{n\kk}}{\varepsilon_{n\kk}-\varepsilon_{n\kk}} = f'_{n\kk}$, we
obtain
\begin{widetext}
  \begin{align}
  \label{eq:chi0}
  \chi_{0}(\rr,\rr') &= \sum_{\kk' \in \mathcal B_{\Ncell}}\sum_{\kk \in \mathcal B_{\Ncell}}\sum_{m=1}^{\infty}\sum_{n=1}^{\infty} \frac{f_{n\kk} - f_{m\kk'}}{\varepsilon_{n\kk} - \varepsilon_{m\kk'}} \cconj{\psi}_{n\kk}(\rr) \psi_{m\kk'}(\rr) \psi_{n\kk}(\rr') \cconj{\psi}_{m\kk'}(\rr') + \frac{|\Omega|}{\Ncell}\frac {D_{\rm loc}(\rr) D_{\rm loc}(\rr')} {D}\\
    \chi_{0}(\qq,\GG,\GG') &= \frac{1}{\Ncell |\Omega|} \sum_{\kk \in \mathcal B_{\Ncell}} \sum_{m=1}^{\infty} \sum_{n=1}^{\infty} \frac{f_{n\kk} - f_{m,\kk-\qq}}{\varepsilon_{n\kk} - \varepsilon_{m,\kk-\qq}} \langle u_{m,\kk-\qq} |e^{-i\GG\cdot\rr}u_{n\kk} \rangle_{\Omega} \langle  u_{n\kk}|e^{i\GG'\cdot\rr} u_{m,\kk-\qq}\rangle_{\Omega}\\
    &+ \delta_{\qq\vec{0}} \frac{1}{D} \langle e^{i\GG\cdot\rr}|D_{\rm loc} \rangle_{\Omega}
			\langle D_{\rm loc}|e^{i\GG'\cdot\rr}\rangle_{\Omega}.\notag
  \end{align}
\end{widetext}
In both these expressions, the second term, arising from the variable
Fermi level, imposes charge neutrality:
$\int_{\Omega_{\Ncell}} \delta\rho(\rr) \D\rr = 0$ for all $\delta V$.

For $\vec{q} \neq \vec{0}$, we can take the
thermodynamic limit \mbox{$\Ncell \to \infty$}
and obtain the usual Adler-Wiser formula
\begin{equation}
\begin{aligned}
  &\lim_{\Ncell\to \infty}\chi_{0}(\qq,\GG,\GG')\\
  &=\frac{1}{(2\pi)^{3}} \int_{k \in \mathcal B} \sum_{n=1}^{\infty} \sum_{m=1}^{\infty} \frac{f_{n\kk} - f_{m,\kk-\qq}}{\varepsilon_{n\kk} - \varepsilon_{m,\kk-\qq}} A_{mn\kk}(\qq,\GG,\GG') \D\kk.
\end{aligned}
\end{equation}
with matrix elements
\begin{equation}
\begin{aligned}
  &A_{mn\kk}(\qq,\GG,\GG')\\
  &\hspace{2em}=\langle  u_{m,\kk-\qq}|e^{-i\GG\cdot\rr}u_{n\kk} \rangle_{\Omega} \langle  u_{n\kk}|e^{i\GG'r}u_{m,\kk-\qq} \rangle_{\Omega}.
\end{aligned}
\end{equation}
Of particular interest is the long-wavelength limit \mbox{$\qq\to \vec{0}$}, $\GG=\vec{0}$, $\GG'=\vec{0}$, in which
case $A_{mn\kk} = \delta_{mn}$ and
\begin{equation}
\begin{aligned}
    \lim_{\qq \to 0} \lim_{\Ncell\to \infty}\chi_{0}(\qq,\vec{0},\vec{0})
  &=\frac{1}{(2\pi)^{3}} \int_{\mathcal B} \sum_{n=1}^{\infty} f'_{n,\kk} \D\kk\\
  &= -\lim_{\Ncell \to \infty} D,
\end{aligned}
\end{equation}
the negative of the density of states at the Fermi level of
this system. In the limit of zero temperature, $f'_{n,\kk}$ localizes
on the Fermi surface, and $D$ converges to a
finite negative value if the system is conducting, and to zero if it
is not. For insulating and semiconducting systems, expanding
$u_{m,\kk+\qq}$ to first order (the ``$\vec{k}\cdot \vec{p}$
perturbation theory''), we obtain that for $\qq$ small, $\chi_{0}(\qq,\vec{0},\vec{0})$ is of the order of $|\qq|^{2}$ and both
$\chi_{0}(\qq,\GG,\vec{0})$ and $\chi_{0}(\qq,\GG',\vec{0})$ are of order of $|\qq|$ for
$\GG, \GG' \neq \vec{0}$.

This has strong consequences for the eigenvalues of the operator
$\epsadj = 1 - \chi_{0} K$. Indeed, in the RPA,
\begin{align}
  \epsadj(\qq,\GG,\GG') = \delta_{\GG\GG'} -  \chi_{0}(\qq,\GG,\GG')\frac{4\pi}{|\qq+\GG'|^{2}}
\end{align}
has the same eigenvalues as the symmetrized operator
\begin{equation}
\begin{aligned}
  (1 - &\vc^{1/2} \chi_{0} \vc^{1/2})(\qq,\GG,\GG')\\
  &= \delta_{\GG\GG'} -4\pi \frac{1}{|\qq+\GG|} \chi_{0}(\qq,\GG,\GG') \frac 1 {{|\qq+\GG'|}}.
\end{aligned}
\end{equation}
It follows from the considerations above that this operator diverges
as $1/|\qq|^{2}$ for $\qq$ small enough in the case of metals or systems
at finite temperature, but not for insulators and semiconductors (see
\citet{cances2010dielectric} as well as \citet{levitt2020screening} for a careful
analysis). When computing for instance on a supercell
$L \times 1 \times 1$ extended in one dimension, the lowest values of
$\qq$ allowed in $\mathcal B_{L}$ are of order $1/L$. The result is that
the condition number $\kappa$ of $\epsadj$ grows like $1/L^{2}$
for metals: this is the source of charge sloshing. For insulators and
semiconductors, no such divergence is present and the 
condition number is independent of the system size. In this case,
\begin{align}
    \varepsilon_{r} = \lim_{\qq\to \vec{0}}\frac{1}{\varepsilon^{-1}(\qq,\vec{0},\vec{0})} > 1
\end{align}
is the macroscopic dielectric constant (relative permittivity). Since
the matrix element $\varepsilon^{-1}(\qq,\vec{0},\vec{0})$ is an upper bound on the
lowest eigenvalue of $\varepsilon^{-1}$, it follows that $\kappa$ is
at least higher than the dielectric constant. Therefore, $\kappa$ can
be rather large for extended systems even if they are not conducting.

\subsection{Neglect of local field effects and dielectric functions}
\label{sec:dielectric_functions}
Since bad convergence is linked to long-wavelength modes, it is useful
to ignore lattice-scale details (``neglect of local field effects'').
Under this approximation, the operator $\chi_{0}$ is modeled as a
translation-invariant operator characterized by its Fourier multiplier
$\chi_{0}(\qq)$. Under the RPA, the dielectric operator is also translation-invariant,
with
\begin{align}
  \varepsilon(\qq) = 1 - \frac{4\pi  \chi_{0}(\qq)}{|\qq|^{2}}
\end{align}
This relatively crude approximation is still able to reproduce at a
qualitative level the difference between metals, semiconductors and
insulators.

For metals,
$\lim_{\qq \to \vec{0}} \chi_{0}^\text{metal}(\qq) = -D$
for $\qq$ small, where $D$ is the density of states per unit volume at
the Fermi level. It follows that an appropriate approximation of
$\varepsilon$ for $\qq$ small is
\begin{align}
  \label{eq:eps_met}
  \varepsilon^\text{metal}(\qq) = \frac{4\pi D + |\qq|^{2}}{|\qq|^{2}}.
\end{align}
Note that this result is the same as that obtained in the Thomas-Fermi theory
of the electron gas. Indeed, in Thomas-Fermi theory, the
potential-to-density mapping is given by $\rho = C_{\rm TF}
(\varepsilon_{F} - V)^{3/2}_{+}$ where $C_{\rm TF}$ is a universal
constant, and therefore
\begin{align}
  \label{eq:chi0_TF}
  \chi_{0}^{\rm TF}(\rr,\rr') = -\frac 3 2 C_{\rm TF}^{\frac 2 3} \rho(r)^{\frac 1 3} \delta(\rr-\rr').
\end{align}
In the case of the homogeneous electron gas where $\rho$ is constant, we recover
\eqref{eq:eps_met}, with
\begin{align}
  \label{eq:DTF}
  D^{\rm TF} = \frac 3 2 C_{\rm TF}^{\frac 2 3} \rho^{\frac 1 3}.
\end{align}
The equation \eqref{eq:eps_met} is also consistent with the small-$\qq$
behavior of the dielectric function derived from the exact
independent-particle susceptibility of the free electron gas (the
Lindhard formula).

For insulators or semiconductors, as we have seen above, a more
appropriate approximation of $\chi_{0}(\qq)$ for $\qq$ small is
$- \qq^{T} \sigma_{0} \qq$, where $\sigma_{0}$ is a unitless
material-dependent symmetric positive definite matrix, with the
physical meaning of a polarizability per unit volume. The dielectric
function is then
\begin{align}
\varepsilon(\qq) = 1+4\pi \frac{\qq^{T} \sigma_{0} \qq}{|\qq|^{2}}
  \label{eq:eps_ins}
\end{align}
In the isotropic case where $\sigma_{0}$ is a scalar,
$\varepsilon_{r} = \varepsilon(\vec{0}) = 1+4\pi \sigma_{0}$ is the
dielectric constant of the material.

\begin{figure}[h!]
  \centering
  \includegraphics[width=0.9\columnwidth]{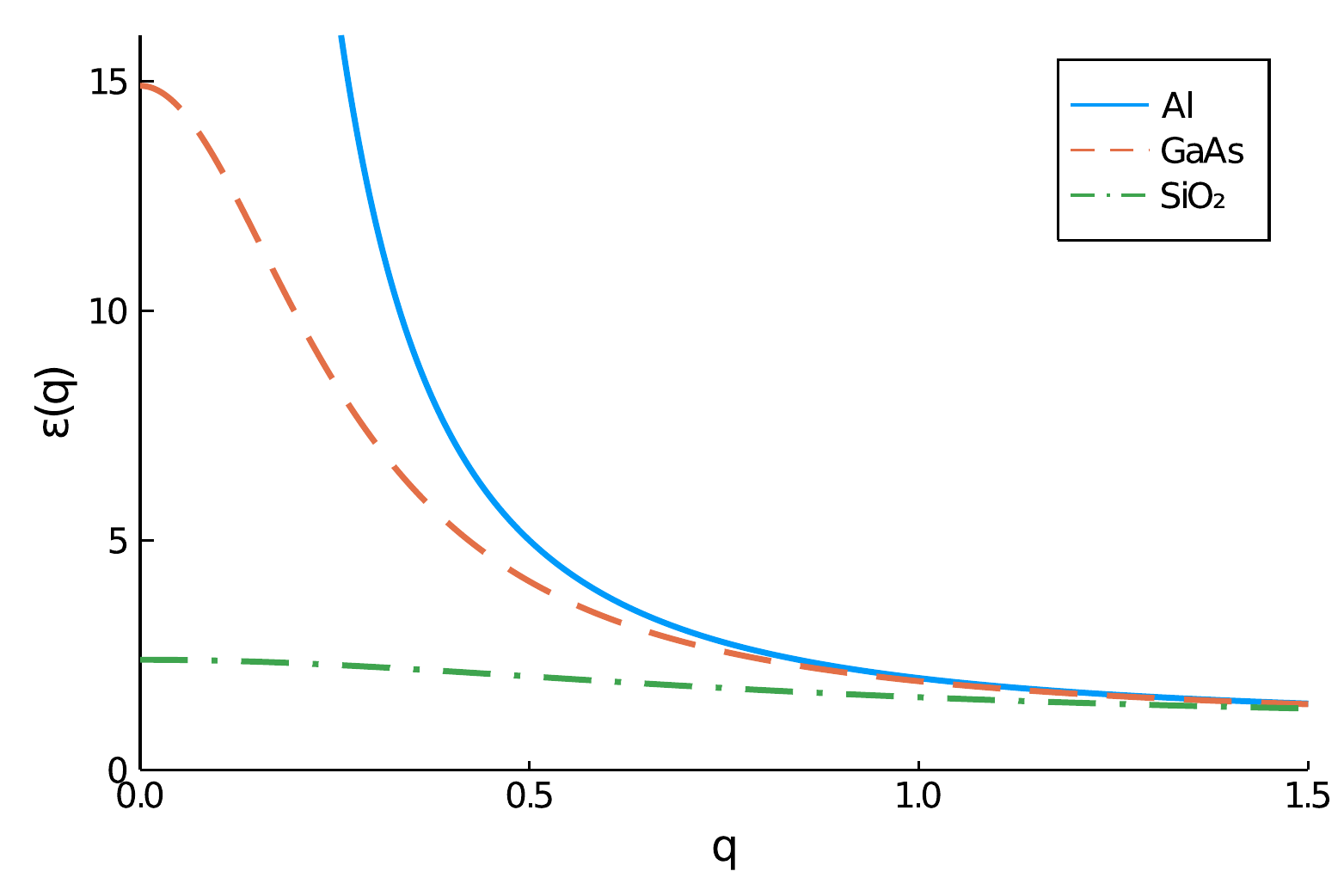}
  \caption{
  Approximate dielectric functions of aluminium, gallium arsenide and silica.
  }
  \label{fig:dielectric_functions}
\end{figure}

The concept of a translation-invariant dielectric function is only
meaningful at long wavelengths (small $\qq$), where local field effects
can be ignored. At larger $\qq$ one can simply extrapolate to a
reasonable function. In Figure \ref{fig:dielectric_functions} we plot
approximate dielectric functions for three bulk materials: aluminium
(\ce{Al}), a metal, gallium arsenide (\ce{GaAs}), a semiconductor, and
silica (\ce{SiO2}), an insulator. For the metal, we used the formula
\eqref{eq:eps_met} above, with $4\pi D = 1$. For the
semiconductor and the insulator, we used the empirical formula
\begin{equation}
\begin{aligned}
  \chi_{0}^\text{dielectric}(\qq) &=-\frac{(\varepsilon_{r} - 1) |\qq|^{2}}{4\pi\left(1 + (\varepsilon_{r}-1)\frac{|\qq|^{2}}{k_\text{TF}^{2}}\right)}\\
  \varepsilon^\text{dielectric}(\qq) &= \frac{\varepsilon_{r} + (\varepsilon_{r}-1) \frac{|\qq|^{2}}{k_\text{TF}^{2}}}{1 + (\varepsilon_{r} - 1) \frac{|\qq|^{2}}{k_\text{TF}^{2}}}
\end{aligned}
	\label{eqn:DielectricModel}
\end{equation}
This is a two-parameter simplification of the Resta model \cite{resta1977thomas} chosen to
reproduce the correct behavior at zero and
at infinity. We used for $\varepsilon_{r}$ the macroscopic isotropic electronic dielectric
constants ($2.4$ for \ce{SiO2} and $14$ for \ce{GaAs}), and set
$k_\text{TF} = 1$.

\subsection{Preconditioners for homogeneous systems}
\label{sec:prechomo}
Several preconditioners have been
proposed in the literature. The simplest class is that of homogeneous
preconditioners, which simply act as multipliers in reciprocal space.
The first such preconditioner is Kerker's scheme \cite{Kerker1981},
derived using
Thomas-Fermi theory:
\begin{align}
  \label{eq:kerker_precon}
  P_{\rm Kerker}^{-1}(\qq) = \frac{1}{\varepsilon^{\rm metal}(\qq)} = \frac{|\qq|^{2}}{k_{\rm TF}^{2} + |\qq|^{2}},
\end{align}
where $k_{\rm TF}$ is the Thomas-Fermi screening wavevector. This is exactly the metallic dielectric
function \eqref{eq:eps_met}, with $k_{\rm TF} = \sqrt{4\pi D}$.

\newcommand{\dielectric}{dielectric\xspace}
\newcommand{\Dielectric}{Dielectric\xspace} This preconditioner is
appropriate for metals but not for insulators and semiconductors, and
therefore the Kerker preconditioner has been modified to accommodate
different types of dielectric behavior, for instance by the truncated
Kerker scheme
$P_{\rm trunc}(\qq) = P_{\rm Kerker}(\max(|\qq|,q_{\rm min}))$ for some
$q_{\rm min}$, or by using the Resta model as preconditioner
\cite{kresse1996efficient, zhou2018applicability,kumar2020preconditioning}. We use a \textit{\dielectric
  preconditioner} based on the dielectric model prescribed in
\eqref{eqn:DielectricModel}:
\begin{equation}
  \label{eq:dielectric_precon}
  P_\text{\dielectric}^{-1}(\qq) = \frac 1 {\varepsilon^\text{dielectric}(\qq)}
    = \frac{ 1+\left(\varepsilon_r-1 \right) \frac{|\qq|^2}{k_\text{TF}^2}}
    {\varepsilon_r +\left(\varepsilon_r-1 \right) \frac{|\qq|^2}{k_\text{TF}^2}}.
\end{equation}
When a preconditioner is chosen, the condition number of the
iterative scheme is equal to the ratio of largest to lowest
eigenvalues of $P^{-1} \varepsilon^{\dagger}$. It follows that the
condition number $\kappa$ of $P^{-1} \varepsilon^{\dagger}$ is small
(and therefore convergence is rapid) when (a) $\varepsilon$ is well-approximated by a
homogeneous dielectric function (and therefore the system is homogeneous
enough), and (b) the dielectric function is well-approximated by the
selected preconditioner (and therefore the parameters are
selected appropriately). Selecting the wrong preconditioner risks
damping too little or too much the long-wavelength modes, resulting in
slow convergence.

\section{Computational results on homogeneous systems}
\label{sec:comphomo}

\newcommand{\nc}{n.c.}        %
\newcommand{\fmtk}{\color{orange!85!black}}  %
\newcommand{\andersoncomment}[1]{
\begin{minipage}{#1}
	\vspace{0.1em}
	\begin{flushleft}
        \footnotesize \textsuperscript{a}System exhibits strong non-linear effects leading to issues with Anderson acceleration.
	\end{flushleft}
\end{minipage}
}
\begin{table*}
    \centering
    \begin{tabular}{l@{\extracolsep{1em}}l  %
                    @{\extracolsep{4em}}r@{\extracolsep{1em}}r@{\extracolsep{3.5em}} %
                    r@{\extracolsep{1em}}r %
                    *{2}{@{\extracolsep{2em}}r@{\extracolsep{1em}}r}}  %
        \hline \hline
        & & \multicolumn{2}{c}{None}
        & \multicolumn{2}{c}{\Dielectric}
        & \multicolumn{2}{c}{\Dielectric}
        & \multicolumn{2}{c}{Kerker} \\
        & & &
        & \multicolumn{2}{c}{(\ce{SiO2})}
        & \multicolumn{2}{c}{(GaAs)}
        & &\\
        & $\mathcal{N}$  & it & $\kappa$ & it & $\kappa$ & it & $\kappa$ & it & $\kappa$ \\\hline
        \ce{SiO2} & 21 & \fmtk{13} & \fmtk{3.2} & \fmtk{13} & \fmtk{3.5} & \fmtk{22} & \fmtk{10.2} & 33  & 103.1 \\
                  & 39 & \fmtk{12} & \fmtk{3.2} & \fmtk{16} & \fmtk{3.5} & \fmtk{22} & \fmtk{10.7} & \nc & 351.6 \\\hline
                  GaAs\textsuperscript{a} & 20 & \fmtk{\nc} & \fmtk{14.7} & \fmtk{\nc} & \fmtk{9.3} & \fmtk{10} & \fmtk{2.4} & 23 & 31.2 \\
                       & 40 & \fmtk{\nc} & \fmtk{18.0} & \fmtk{\nc} & \fmtk{11.3} & \fmtk{10} & \fmtk{2.7} & 38 & 100.9 \\\hline
        Al & 20 & 29  & 58.1 & 22 & 30.5 & 15 & 7.2 & \fmtk{10} & \fmtk{3.9} \\
           & 40 & \nc & 261.6 & \nc & 141.3 & 24 & 27.8 & \fmtk{10} & \fmtk{2.5} \\
        \hline \hline
    \end{tabular}
    \andersoncomment{0.64\textwidth}
    \caption{
        Number of iterations and condition number $\kappa$
        for several homogeneous materials and preconditioners.
        $\mathcal{N}$ denotes the number of repeats of the conventional
        unit cell and ``\nc'' indicates that the SCF has not converged
        after $50$ iterations.
        The cases where $\kappa$ increases by less
        than a factor of two when the system size is doubled
        are colored.
	}
	\label{tab:bulk}
\end{table*}

To compare different strategies of preconditioning numerically we will
consider six testcases in the form of elongated supercells of
silica~(\ce{SiO2}), gallium arsenide~(\ce{GaAs}) and
aluminium~(\ce{Al}) as simple examples of a bulk insulator,
semiconductor and metal respectively.

To generate these test systems we repeated the conventional cubic (for
\ce{GaAs} and \ce{Al}) or trigonal (for \ce{SiO2}) unit cells of the
respective materials roughly 20 or 40 times along the $(001)$
surface. All calculations have been performed in the
density-functional toolkit~(DFTK)~\cite{DFTK}, in a plane-wave
basis, using the
Perdew-Burke-Ernzerhof~(PBE) exchange-correlation
functionals~\cite{Perdew1996} as implemented in the
libxc~\cite{Lehtola2018} library, and Godecker-Teter-Hutter
pseudopotentials~\cite{Goedecker1996}. We used a kinetic energy cutoff
of \unit[20]{Hartree} and Monkhorst-Pack grids with a maximal spacing
of \unit[0.3]{Bohr$^{-1}$}. For aluminium a Gaussian
smearing scheme with width of \unit[0.001]{Hartree} was used. We chose
tight convergence tolerances for the diagonalization during the SCF
procedure to avoid any slowdown due to inaccurately represented bands. To avoid symmetry effects which would result in faster
convergence not representative of larger-scale computations, we
slightly perturbed the atomic positions and the initial guess density.
We refer to our github repository of supporting
information~\cite{reproducers} for full computational details,
including instructions on how to reproduce all the results in this
paper.

We show in Table \ref{tab:bulk} the results of the SCF procedure with
Anderson acceleration (history size $m=10$) in four cases: without
preconditioning ($P=1$), with the dielectric preconditioner
\eqref{eq:dielectric_precon} (using the dielectric constants of
\ce{SiO2} and \ce{GaAs}), and with the Kerker preconditioner. In all
cases we took $k_{\rm TF}=1$. In order to avoid selecting good
stepsizes manually, we estimated the lowest and largest eigenvalues
$\lambda_{\min}$ and $\lambda_{\max}$ of $P^{-1} \varepsilon^\dagger$
at convergence, and used these to determine the optimal stepsize from
\eqref{eq:optimal_step}. The eigenvalues were determined iteratively
using the Arnoldi method. At each step, the action of
$\varepsilon^\dagger = 1 - K \chi_{0}$ on a vector was computed by
solving the Sternheimer equations. A similar method was used
by Wilson and coworkers~\cite{Wilson2008,Wilson2009}
to approximate the dielectric operator.
For technical reasons, we used the Teter reparametrization of the
local-density approximation (LDA)~\cite{Goedecker1996}
in the exchange-correlation kernel for the
computation of $K$ instead of the PBE functional used in the SCF
computations themselves.

From the eigenvalues we also computed the condition number
$\kappa = \frac {\lambda_{\max}}{\lambda_{\min}}$. This gives a simple intrinsic
measure of the quality of the preconditioner in the linear regime,
free from the noise associated with the Anderson method (nonlinear
effects, stepsize, history length, implementation details). Since the
eigenvalues can be slow to converge, we only report approximate
values. Using a crude bound based on the residuals obtained
in our Arnoldi scheme we estimate
the error in the values for $\kappa$ to be less than \unit[13]{$\%$}
over all cases considered in this work.

\begin{figure}
	\centering
	\includegraphics[width=0.9\columnwidth]{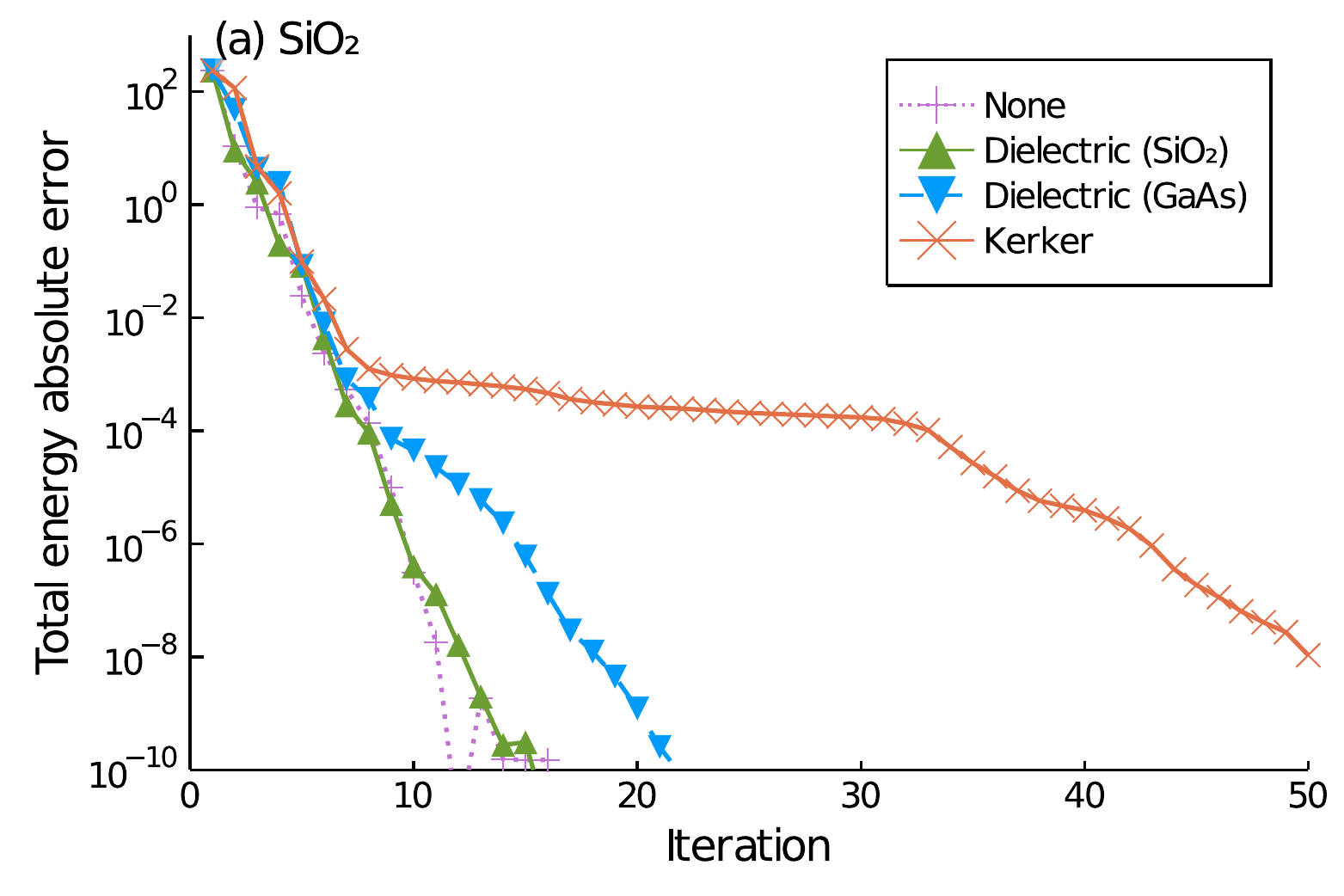}
	\includegraphics[width=0.9\columnwidth]{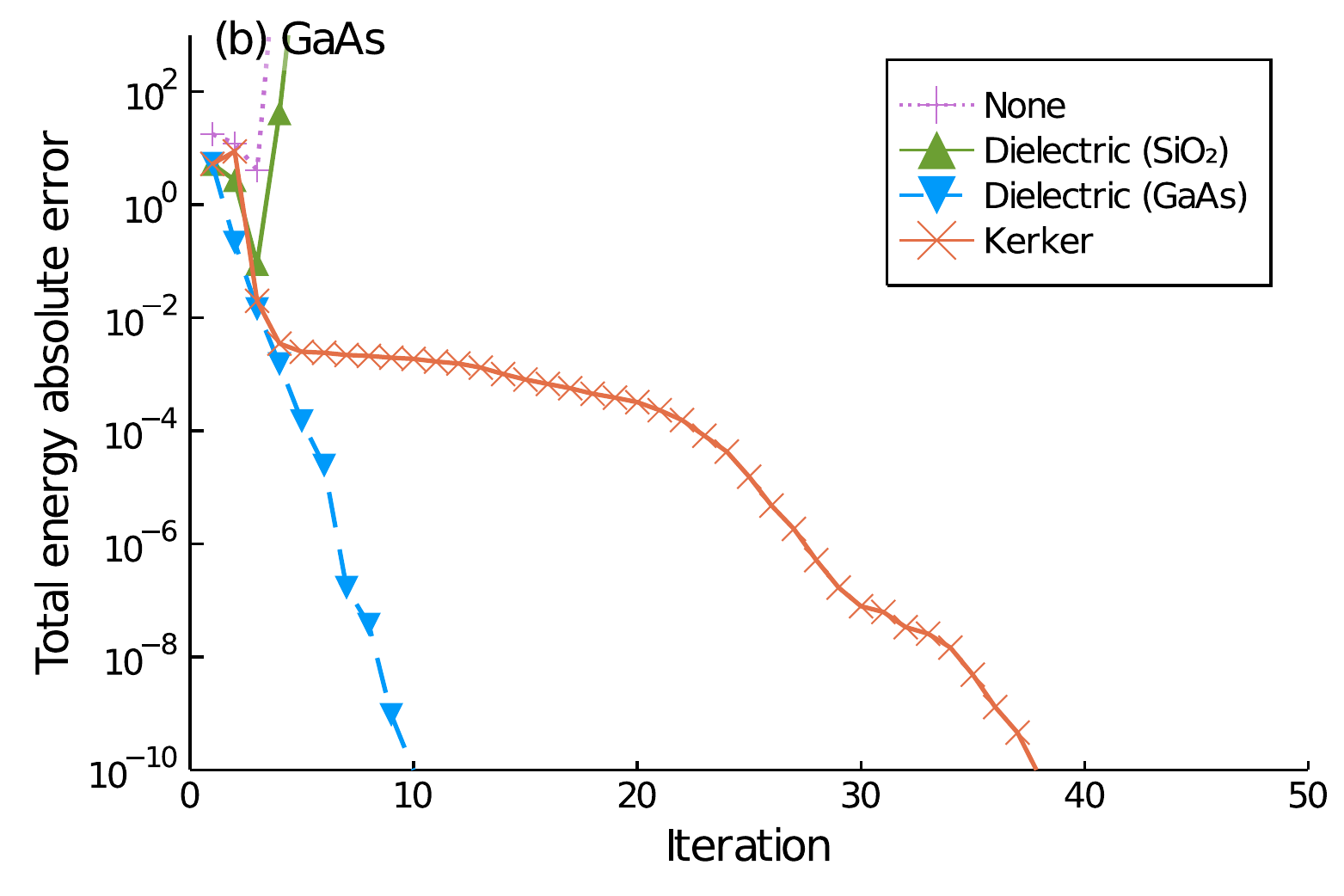}
	\includegraphics[width=0.9\columnwidth]{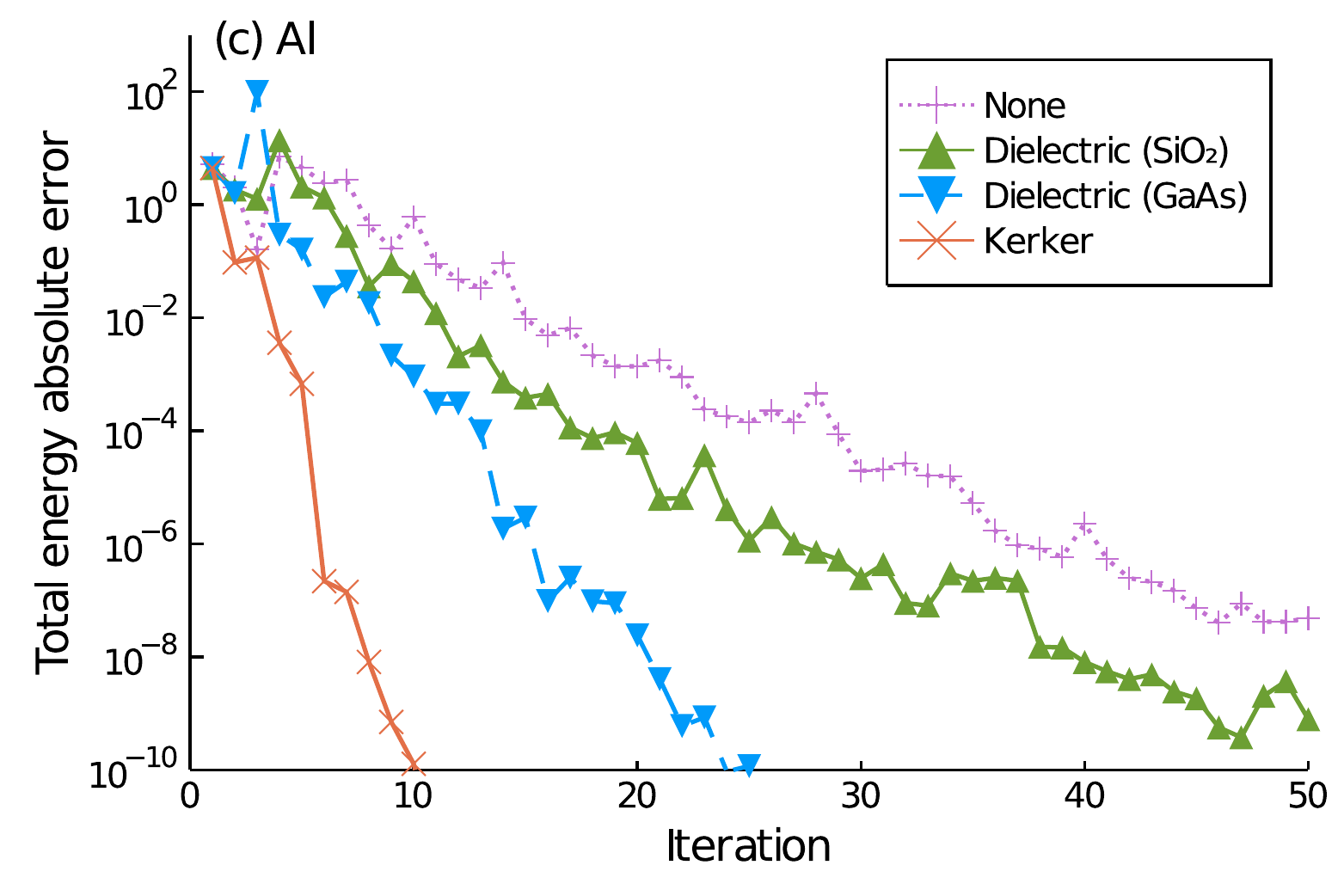}
	\caption{Convergence plots for bulk systems.
        From top to bottom the plots show (a) 39 repeats of silica,
        (b) 40 repeats of gallium arsenide and (c) 40 repeats of aluminium.
        For \ce{GaAs} the iterations for no preconditioner
        and for the dielectric (\ce{SiO2}) preconditioner start to diverge
        after 3 iterations.
    }
	\label{fig:bulk}
\end{figure}

Both the number of iterations required to converge the SCF
to an energy error of $10^{-10}$ as well as the condition number $\kappa$
are summarized in Table~\ref{tab:bulk}. The corresponding plots are
shown in Figure~\ref{fig:bulk}.

The combinations of systems and preconditioners for which there is no
significant growth in the number of iterations and 
condition numbers are highlighted in orange color in Table~\ref{tab:bulk}. As
predicted from the theory, Kerker preconditioning leads to system size
independence for the metal \ce{Al}, and no preconditioning or
dielectric preconditioning for the insulator
\ce{SiO2} or the semiconductor \ce{GaAs}. In the other cases, the
condition number grows quadratically with the system size.
On semiconductors, even if the unpreconditioned condition
number is independent on system size, it can still be rather large (on the
order of the dielectric constant) and therefore benefit from
appropriate preconditioning.

On the diagonal in Table~\ref{tab:bulk}, using a preconditioner
adapted to the system at hand leads to a condition number between 2
and 4, and convergence in less than 20 iterations. Using the wrong
preconditioner leads to degraded convergence. On \ce{SiO2} and
\ce{Al}, convergence is eventually achieved even for inadequate
preconditioners, albeit slowly. On \ce{GaAs} however, the interference
of strong nonlinear effects on the Anderson scheme leads to a quick
divergence (see discussion in Section~\ref{sec:Anderson}).

\begin{figure}
    \centering
    \includegraphics[width=0.9\columnwidth]{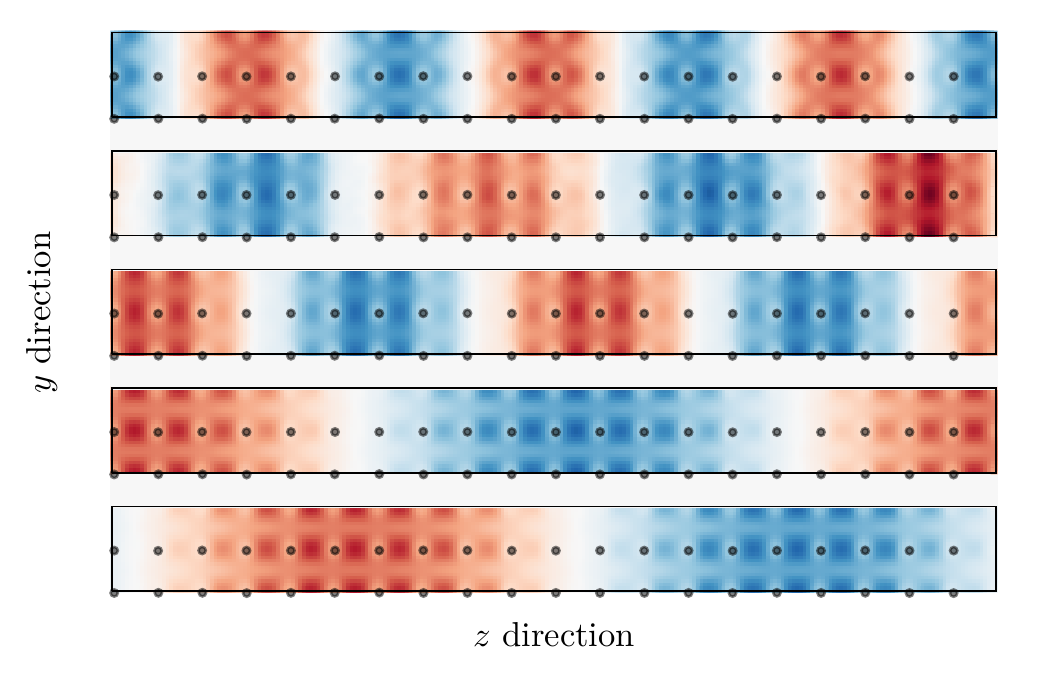}
    \caption{The five most dominant eigenmodes
        of the dielectric adjoint operator $\varepsilon^\dagger$
        for bulk aluminium with 40 repeats.
        The modes are plotted in real space,
        averaged over the $x$ direction,
        corresponding approximately (from top to bottom)
        to eigenvalues $24$, $46$, $47$, $261$ and $261$.
        The modes display typical
        charge-sloshing behavior,
        moving electron density between extremal parts
        of the cell.
    }
    \label{fig:modesAluminium}
  \end{figure}

The long-wavelength charge sloshing is responsible for slow
convergence, as illustrated Figure~\ref{fig:modesAluminium} on the
largest eigenmodes of $\epsadj$ in the case of unpreconditioned
\ce{Al}. Being eigenmodes of the almost periodic operator $\epsadj$, these
modes are close to Bloch waves $e^{i\qq\cdot \rr} u(\rr)$ with $\qq$
consistent with periodic boundary conditions on the computational
domain and $u$ lattice-periodic. The largest eigenvalues are
associated with small $\qq$.

\section{Preconditioners for heterogeneous systems}
\label{sec:prechetero}
As we have seen, for simple homogeneous systems a
translation-invariant approximation of $\chi_{0}$ and therefore
$\varepsilon^{-1}$ is appropriate.
For more complex systems
(for instance, periodic materials with large unit cells, clusters and surfaces),
these are very often
unsuccessful in significantly reducing the number of iterations, and
several alternative preconditioners have been proposed.

The scheme in \citet{raczkowski2001thomas} employs
a modified Thomas-Fermi-von~Weizsäcker~(TFW) model to derive an
inexpensive approximation to the dielectric operator.
However, the TFW model, like the simpler
Thomas-Fermi model in \eqref{eq:chi0_TF}, predicts metallic-like full screening (see
\citet{cances2011local} for a detailed analysis), and
is therefore unable to account for the difference in dielectric behavior between
insulators, semiconductors and metals. In practice, we observed the convergence
with this preconditioner to strongly degrade with system
size for semiconductors and (especially) insulators, exactly as the Kerker scheme.
A slightly generalized version of this approach has been proposed by
\citet{hasnip2015auxiliary}, where the Kohn-Sham problem is
dynamically mapped to a related auxiliary problem,
which is solved self-consistently at
each iteration in order to determine the preconditioned density for
the next iteration. The ``extrapolar method'' of
\citet{anglade2008preconditioning}, building on the earlier HIJ method
\cite{ho1982dielectric}, is based on a full computation and inversion
of the dielectric operator on a reduced system. It performs well,
but is very costly, scaling like the fourth power of the number of
electrons. The ``elliptic preconditioner'' of \citet{lin2013elliptic}
uses a model elliptic equation with variable coefficients as a
dielectric model, approximating $\varepsilon^{-1}$ by the
solution of the equation
\begin{align}
  \left[\nabla \cdot \big(a(\rr) \nabla\big) + 4\pi b(\rr)\right] (\varepsilon^{-1} V)
  = -\Laplace V.
\end{align}
In the case of constant coefficients $a$ and $b$, this method is able
to reproduce the dielectric behavior of metals \eqref{eq:eps_met} by
taking $a = 1$, $b = D$ and of insulators and semiconductors
\eqref{eq:eps_ins} by taking $b = 0$ and $a$ equal to the dielectric
constant. By manually constructing appropriate switching functions for
$a$ and $b$, they
were able to demonstrate good performance on hybrid systems consisting
of a metallic and a vacuum region. However, this method requires
manual tuning for the system at hand, in a fashion that is hard to
automatize.

In this work, we propose to build an inhomogeneous preconditioner by
approximating the independent-particle susceptibility $\chi_{0}$
instead of the dielectric operator $\varepsilon$.
Consider an arbitrary finite system (not necessarily periodic) with
orthonormal eigenfunctions $\varphi_{i}$, eigenvalues
$\varepsilon_{i}$, Fermi level $\varepsilon_{F}$, occupations $f_{i} =
f\left(\frac{\varepsilon_{i}-\varepsilon_{F}}{T}\right)$ and temperature $T$. Then, following the computations in Section~\ref{sec:chi0_deriv}, $\chi_{0}$ can be
written as
\begin{equation}
  \begin{aligned}
    \chi_{0}(\rr,\rr') &= \sum_{i=1}^{\infty} \sum_{j=1}^{\infty}
    \frac{f_{i}-f_{j}}{\varepsilon_{i}-\varepsilon_{j}}
    \cconj{\varphi}_{i}(\rr) \varphi_{j}(\rr) \varphi_{i}(\rr') \cconj{\varphi}_{j}(\rr') \\&+
    \frac{D_{\rm loc}(\rr)D_{\rm loc}(\rr')}{D},
  \end{aligned}
  \label{eqn:chi0arbitrary}
\end{equation}
where
\begin{equation}
\label{eqn:ldoscompute}
\begin{aligned}
  D &= -\sum_{i=1}^{\infty} f'_{i}\\
  D_{\rm loc}(\rr) &= -\sum_{i=1}^{\infty} f'_{i} |\varphi_{i}|^{2}(\rr)
\end{aligned}
\end{equation}
are the density of states (not per volume, in contrast to the
computations in \ref{sec:chi0_deriv}) and localized density of states
at the Fermi level.

The second term $\chi_{0}^{(2)}$ in \eqref{eqn:chi0arbitrary} relates
to the global variation of the Fermi level, is fully nonlocal and
easily computable. The first term $\chi_{0}^{(1)}$ with its double sum
over all states is where the complexity of building preconditioners
lies. However, if we are interested only in describing long-range
modes, we can write
\begin{equation}
  \label{eq:chi0_approx}
  \begin{aligned}
    \int \chi_0^{(1)}(\rr,\rr') V(\rr') \D\rr'
    &\approx V(\rr) \int \chi_{0}^{(1)}(\rr,\rr') \D\rr' \\
    &=V(\rr) \sum_{i=1}^{\infty}\sum_{j=1}^{\infty} \frac{f_{i}-f_{j}}{\varepsilon_{i}-\varepsilon_{j}} \cconj{\varphi}_{i}(\rr) \varphi_{j}(\rr) \delta_{ij}\\
    &= -D_{\rm loc}(\rr) V(\rr)
  \end{aligned}
\end{equation}
where we have assumed that $\rr' \mapsto \chi_{0}^{(1)}(\rr,\rr')$ is localized
around $\rr$ at a scale smaller than the characteristic scale of
variation of $V$.
This process corresponds to condensing all of $\chi_{0}^{(1)}(\rr,\rr')$
on its diagonal (similar to the ``mass lumping''
used in the finite element community).
\begin{algorithm}[H]
    \caption{The LDOS preconditioner $P^\text{LDOS}$
        within a preconditioned SCF scheme \eqref{eqn:Preconditioning}}
  \label{alg:ldos}
   \begin{algorithmic}[1]
       \Require{$\delta F = F(V(\rho_n)) - \rho_n$,
           current $\{\varepsilon_{n\kk}\}$ and $\{\psi_{n\kk}\}$}
       \Ensure{$\delta\rho$
           to construct $\rho_{n+1} = \rho_n + \alpha \delta\rho$}\\[]
       \Function{ldos\_preconditioner}{$\delta F, \{\varepsilon_{n\kk}\}, \{\psi_{n\kk}\}$}
       \State Compute $D$ and $D_\text{loc}$, see \eqref{eqn:ldoscompute}
       \State Solve $(1 - \chi_0^\text{LDOS} \vc) \,\delta\rho = \delta F$ for $\delta\rho$ using GMRES\cite{Saad2003}
       \State \Return $\delta\rho$
       \EndFunction
   \end{algorithmic}
\end{algorithm}
\noindent Remarkably, this large-scale averaging process
involves only the ``diagonal'' excitations $i=j$ in $\chi_{0}^{(1)}$,
resulting in a tremendous reduction in complexity.

The final expression (and the central result of this paper) is
\begin{equation}
  \chi_{0}^\LDOS(\rr,\rr') = -D_{\rm loc}(\rr) \delta(\rr-\rr') + \frac{D_{\rm loc}(\rr)D_{\rm loc}(\rr')}{D}.
    \label{eqn:chi0ldos}
  \end{equation}
  We use this to build a preconditioner
  \begin{align}
    P^{\LDOS} =
  (1 - \chi_{0}^\LDOS \vc\big).
\end{align}
In our tests, including the exchange-correlation kernel $K_{\rm XC}$
in the preconditioner did not improve it, and therefore we stay within
the RPA.

To apply $P^{-1}$ to a vector, we solve the linear equation
iteratively using the GMRES method~\cite{Saad2003},
see Algorithm \ref{alg:ldos}.
We first compute the local density of states, which is of the same
complexity as computing the density.
At each step of the iterative solver (line 3 of Algorithm \ref{alg:ldos}),
we have to apply $\vc$ and $\chi_{0}^{\LDOS}$ to a test vector.
These are inexpensive operations since they scale linearly with the number of atoms.
More precisely, applying $\chi_{0}^{\LDOS}$ only requires
multiplications in real space and the Coulomb operator
is applied as usual (using Fourier transforms and a multiplication in Fourier space).
Both operations are needed for other steps in DFT computations,
and therefore this method can readily be
implemented in any DFT code, independently of the basis set used.

This preconditioner is completely parameter-free, and can be seen as a
localized version of Kerker's
preconditioning, to which it reduces when $D_{\rm loc}$ is constant.
By using the density of states of the system, it automatically
adjusts the constant $k_{\rm TF}^{2}$ appearing in the Kerker
preconditioner using the properties of the system under consideration.
This systematizes the findings in \citet{zhou2018applicability} where this constant was adapted to
the number of metallic atoms in a hybrid Au-MoS2 system. In contrast to the homogeneous preconditioners discussed
in Section \ref{sec:prechomo}, the LDOS-based approach furthermore does not
assume that local field effects can be neglected, since the inverse of
$P$ is computed iteratively, which is more appropriate to heterogenous
systems.

Since the LDOS is localized on metallic regions, we can expect the approximation
\eqref{eqn:chi0ldos} to be accurate on systems made of simple, free
electron-like metals and vacuum, and we will see in
Section~\ref{sec:comphetero} that this is indeed the case. On regions containing insulators and
semiconductors, $D_{\rm loc}(\rr)$ will be zero, such that
\eqref{eqn:chi0ldos} also performs well for almost inert materials
like \ce{SiO2}, where $\chi_{0} \approx 0$ is an appropriate
approximation. This is in contrast to the preconditioner of \citet{raczkowski2001thomas} based on Thomas-Fermi theory, which
incorrectly predicts a non-zero local density of states proportional
to the cube root of the density (see \eqref{eq:DTF}), and therefore in particular
overscreens insulators. However, on semiconductors like gallium arsenide, our
approximation corresponds to not preconditioning at all, which we saw
in Section \ref{sec:comphomo} yields a rather large condition
number of the order of its dielectric constant. This can be traced to
the fact that the approximation \eqref{eq:chi0_approx} completely
neglects any type of polarizability and only considers the bulk
reaction of free electrons.

\newcommand{\ldosResta}{LDOS+dielectric\xspace}
To account for this and obtain a model which is suitable
for semiconductors as well, we can simply add the susceptibility model of \eqref{eqn:DielectricModel}:
\begin{equation}
  \chi_{0}^\text{\ldosResta} = \chi_{0}^\LDOS + \chi_0^\text{dielectric}.
    \label{eqn:chi0empirical}
\end{equation}
The rationale behind this empirical modification is that adding the
dielectric model corrects for $\chi_{0}^{\LDOS}$ being zero in
semiconducting regions while at the same time not having a large
effect in metallic regions, where the local density of states term of
$\chi_{0}^{\LDOS}$ is anyway dominating. Compared to the pure
$\chi_{0}^\LDOS$ model, which is completely free of empirical
parameters, this ``\ldosResta'' preconditioner introduces the two free
parameters $k_\text{TF}$ and $\varepsilon_r$ from
$\chi_0^\text{dielectric}$. As we will see in the following this does,
however, allow the resulting \ldosResta preconditioner to be much more
effective on mixed systems containing semiconductors. Naturally
further improvements are possible, at the expense of introducing
additional empirical parameters. One aspect we have considered is the
use of a switching function $L(\rr)$ going from $0$ to $1$ inside the
semiconducting region of the system.
This effectively activates
$\chi_{0}^{\text{dielectric}}$ only in the semiconducting region,
resulting in the ``localized'' preconditioner
\begin{equation}
  \label{eq:localized_precond}
  \begin{aligned}
  \chi_{0}^{\text{localized}}(\rr,\rr') &= \chi_{0}^\LDOS(\rr,\rr') \\
  &+ \sqrt{L(\rr)} \chi_0^\text{dielectric}(\rr,\rr') \sqrt{L(\rr')}.
  \end{aligned}
\end{equation}

\section{Computational results on heterogeneous systems}
\label{sec:comphetero}

\begingroup
\squeezetable
\begin{table*}
    \centering
    \begin{tabular}{
            l@{\extracolsep{1em}}l  %
            @{\extracolsep{4em}}r@{\extracolsep{1em}}r@{\extracolsep{3.5em}} %
            r@{\extracolsep{1em}}r %
            *{5}{@{\extracolsep{2em}}r@{\extracolsep{1em}}r}  %
        }
        \hline\hline
        & & \multicolumn{2}{c}{None}
        & \multicolumn{2}{c}{\Dielectric}
        & \multicolumn{2}{c}{Kerker}
        & \multicolumn{2}{c}{LDOS}
        & \multicolumn{2}{c}{LDOS+}
        & \multicolumn{2}{c}{Localized} \\
        & & & & & & & & & & \multicolumn{2}{c}{Dielectric} & & \\
        & $\mathcal{N}$  & it & $\kappa$ & it & $\kappa$ & it & $\kappa$ & it & $\kappa$ & it & $\kappa$ & it & $\kappa$ \\\hline
        \ce{SiO2}+vacuum & 10 & \fmtk{11} & \fmtk{3.3} & \fmtk{26} & \fmtk{19.7} & 50 & 95.7 & \fmtk{11} & \fmtk{3.3} & \fmtk{26} & \fmtk{19.7} &&\\
        & 20 & \fmtk{12} & \fmtk{3.4} & \fmtk{30} & \fmtk{24.4} & \nc & 351.5 & \fmtk{12} & \fmtk{3.4} & \fmtk{30} & \fmtk{21.7} &&\\\hline
        GaAs+vacuum & 10 & \fmtk{17} & \fmtk{13.4} & 18 & 6.2 & 23 & 67.0 & \fmtk{17} & \fmtk{12.4} & \fmtk{18} & \fmtk{10.4} & \fmtk{11} & \fmtk{3.5} \\
        & 20 & \fmtk{20} & \fmtk{15.5} & 22 & 12.9 & \nc & 312.2 & \fmtk{20} & \fmtk{15.5} & \fmtk{22} & \fmtk{12.9} & \fmtk{13} & \fmtk{4.6} \\\hline
        Al+vacuum & 10 & 19 & 51.5 & 24 & 44.3 & 22 & 64.4 & \fmtk{9} & \fmtk{3.7} & \fmtk{16} & \fmtk{10.3} & & \\
        & 20 & 47 & 170.8 & 49 & 168.5 & \nc & 323.9 & \fmtk{9} & \fmtk{3.5} & \fmtk{20} & \fmtk{10.5} & & \\\hline
        GaAs+\ce{SiO2}\textsuperscript{a} & 10 & \fmtk{45} & \fmtk{13.7} & \fmtk{19} & \fmtk{8.9} & 34 & 52.4 & \fmtk{45} & \fmtk{13.4} & \fmtk{19} & \fmtk{8.8} & \fmtk{16} & \fmtk{6.1} \\
        & 20 & \fmtk{\nc} & \fmtk{18.2} & \fmtk{20} & \fmtk{10.2} & \nc & 170.1 & \fmtk{\nc} & \fmtk{18.2} & \fmtk{20} & \fmtk{10.2} & \fmtk{24}& \fmtk{10.3} \\\hline
        Al+\ce{SiO2} & 10 & 43 & 93.1 & 29 & 33.6 & 30 & 50.9 & \fmtk{17} & \fmtk{6.1} & \fmtk{20} & \fmtk{9.2} & & \\
        & 20 & \nc & 316.6 & \nc & 118.4 & \nc & 159.4 & \fmtk{14} & \fmtk{5.4} & \fmtk{20} & \fmtk{10.1} & & \\\hline
        Al+GaAs & 10 & \nc & 144.0 & 24 & 22.4 & 16 & 9.0 & 15 & 7.2 & \fmtk{11} & \fmtk{3.5} & & \\
        & 20 & \nc & 485.0 & 40 & 59.0 & 26 & 28.8 & 26 & 21.4 & \fmtk{13} & \fmtk{5.0} & & \\
        \midrule[1pt]
        Al+GaAs+\ce{SiO2} & 10 & \nc & 149.5 & 34 & 50.4 & 36 & 62.9 & 26 & 21.5 & 19 & 9.0 & 21 & 10.4 \\
        \hline\hline
    \end{tabular}
    \andersoncomment{0.85\textwidth}
    \caption{
        Number of iterations and condition number $\kappa$
        for several heterogeneous materials and preconditioners.
        $\mathcal{N}$ denotes the number of repeats of the conventional
        unit cell and ``\nc'' indicates that the SCF has not converged
        after $50$ iterations.
        The cases where $\kappa$ increases by less
        than a factor of two when the system size is doubled
        are colored.
    }
	\label{tab:mixed}
\end{table*}
\endgroup

In this section we consider a number of interfaces involving
a combination of the materials we used previously (\ce{SiO2}, \ce{GaAs}
and \ce{Al}) as well as vacuum. The computational details are the same as in
Section~\ref{sec:comphomo}, with the exception of the $k$-point mesh, for
which a maximal spacing of \unit[0.15]{Bohr$^{-1}$} was used, and the
initial guess and structures, which were not randomized.

We built the surface systems by stacking $\mathcal{N}$
monolayers of the first material with $\mathcal{N}$ monolayers of the
second material and, for \ce{Al}+\ce{GaAs}+\ce{SiO2}, additionally
$\mathcal{N}$ monolayers of the third material.
For \ce{SiO2} and \ce{GaAs} the $(110)$ surface was employed,
while the $(100)$ surface was used for \ce{Al}.
For cases involving vacuum the vacuum region was taken of the same
width as the material layer.
To avoid the emergence of surface states,
which are highly localized and thus
form another source of instability not of focus here,
all silica surfaces were passivated with hydrogen.
To make the lattices compatible, \ce{Al} and \ce{SiO2} were
compressed or expanded in the appropriate direction to adapt to \ce{GaAs}.
Note that we did not deform \ce{GaAs}, where the band gap is most
susceptible to strain.
We tested the following preconditioners:
\begin{itemize}
\item no preconditioner ($P=1$);
\item the \dielectric preconditioner \eqref{eqn:DielectricModel}, with
$\varepsilon_{r} = 14$, $k_{\text{TF}} = 1$;
\item the Kerker
  preconditioner \eqref{eq:kerker_precon}, with $k_{\text{TF}}=1$;
\item the parameter-free LDOS preconditioner based
  on \eqref{eqn:chi0ldos};
\item the \ldosResta preconditioner based on
  \eqref{eqn:chi0empirical}, with
$\varepsilon_{r} = 14$, $k_{\text{TF}} = 1$;
\item on selected systems, the localized preconditioner based on
  \eqref{eq:localized_precond}, with the same parameters as the
  \ldosResta preconditioner
  and $L(\rr)$ constructed as a linear combination
  of error functions chosen to be $1$ inside the \ce{GaAs} region and $0$ outside;
\item on selected systems, the TFW preconditioner of
  \citet{raczkowski2001thomas}, using the implementation in
  Quantum ESPRESSO~\cite{Giannozzi2009,Giannozzi2017} with the same
  parameters as for the other computations as far as possible.
  For these computations we use Quantum ESPRESSO's default
  damping value of $0.7$, but found results to be similar for
  other choices of damping. In contrast to the DFTK-computed curves,
  where the absolute error in the total energy is used directly to
  estimate the SCF error at each iteration, the Quantum ESPRESSO
  curves show the \texttt{estimated scf accuracy} quoted in the
  Quantum ESPRESSO output (which is the $v_c$-norm of the density
  difference squared).
\end{itemize}

As previously, we show the number of iterations to reach an energy
error of $10^{-10}$ and the condition numbers $\kappa$ in
Table~\ref{tab:mixed}. Cases where $\kappa$ increases by less than a
factor of two as the system size is doubled are again highlighted in
orange. In every system except for \ce{GaAs}+\ce{SiO2},
the SCF procedures converge or are on the way to very slow convergence
when they are stopped after 50 iterations.
In \ce{GaAs}+\ce{SiO2} however, as in bulk
\ce{GaAs}, the interference of strong nonlinear effects on the
Anderson scheme leads to a quick divergence for some preconditioners.

None of the homogeneous strategies are able to perform well on all
surfaces. This is particularly clear in the case of systems containing
aluminium: for instance, on the large \ce{Al}+\ce{SiO2} and \ce{Al}+vacuum
test case, all have condition numbers in the hundreds,
with the SCF converging very slowly (in around 50 iterations or more).
Even on nonmetallic systems, the homogeneous preconditioners are not able to achieve condition
numbers less than $10$ on the large testcases, with the exception of
the unpreconditioned \ce{SiO2} surface.%

By contrast, our parameter-free LDOS-based scheme performs fairly
well, achieving condition numbers less than 10 for all
systems not containing the semiconductor \ce{GaAs}, with associated
quick convergence of the SCF. Even for the systems containing
\ce{GaAs}, condition numbers are reasonable, never exceeding 25. The
convergence is accordingly relatively quick, with the exception of
the \ce{GaAs}+\ce{SiO2} system mentioned above.

On systems containing \ce{GaAs}, we tested the \ldosResta and the
localized preconditioner. In \ce{Al}+\ce{GaAs}, adding the
dielectric susceptibility model results in a low condition number,
suggesting that it does not significantly hamper the good model in the
metallic region. In the \ce{GaAs} surface (where the LDOS is constant zero),
plainly adding the susceptibility model does not significantly improve things.
However, localizing it on the \ce{GaAs} results in a low condition
number. The improvement is not as marked on the \ce{GaAs}+\ce{SiO2},
where surface effects might play a larger role. Finally, the larger
\ce{Al}+\ce{GaAs}+\ce{SiO2} is adequately preconditioned with the
\ldosResta model, without need of localizing it.

\begin{figure}
    \centering
    \includegraphics[width=0.9\columnwidth]{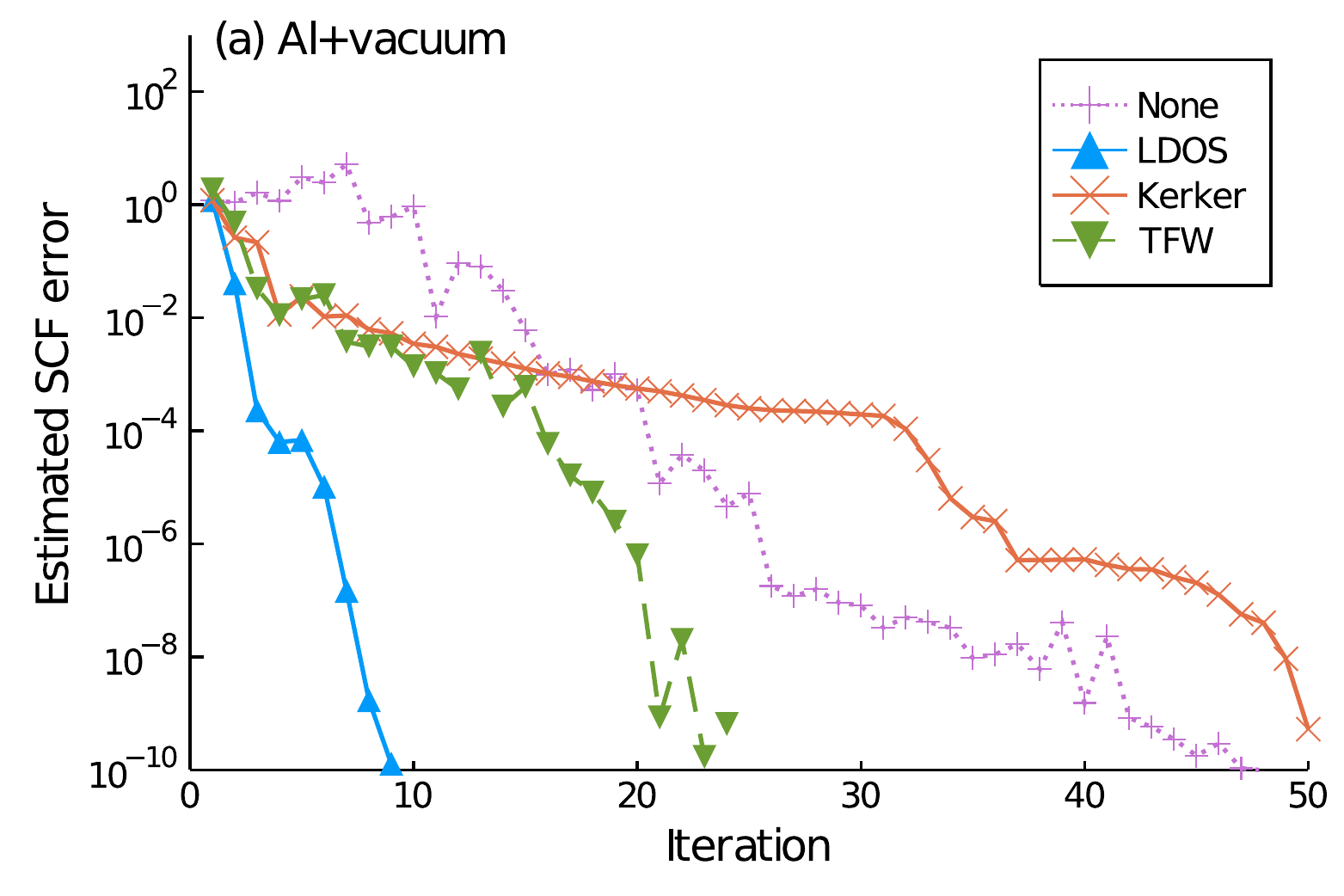}
    \includegraphics[width=0.9\columnwidth]{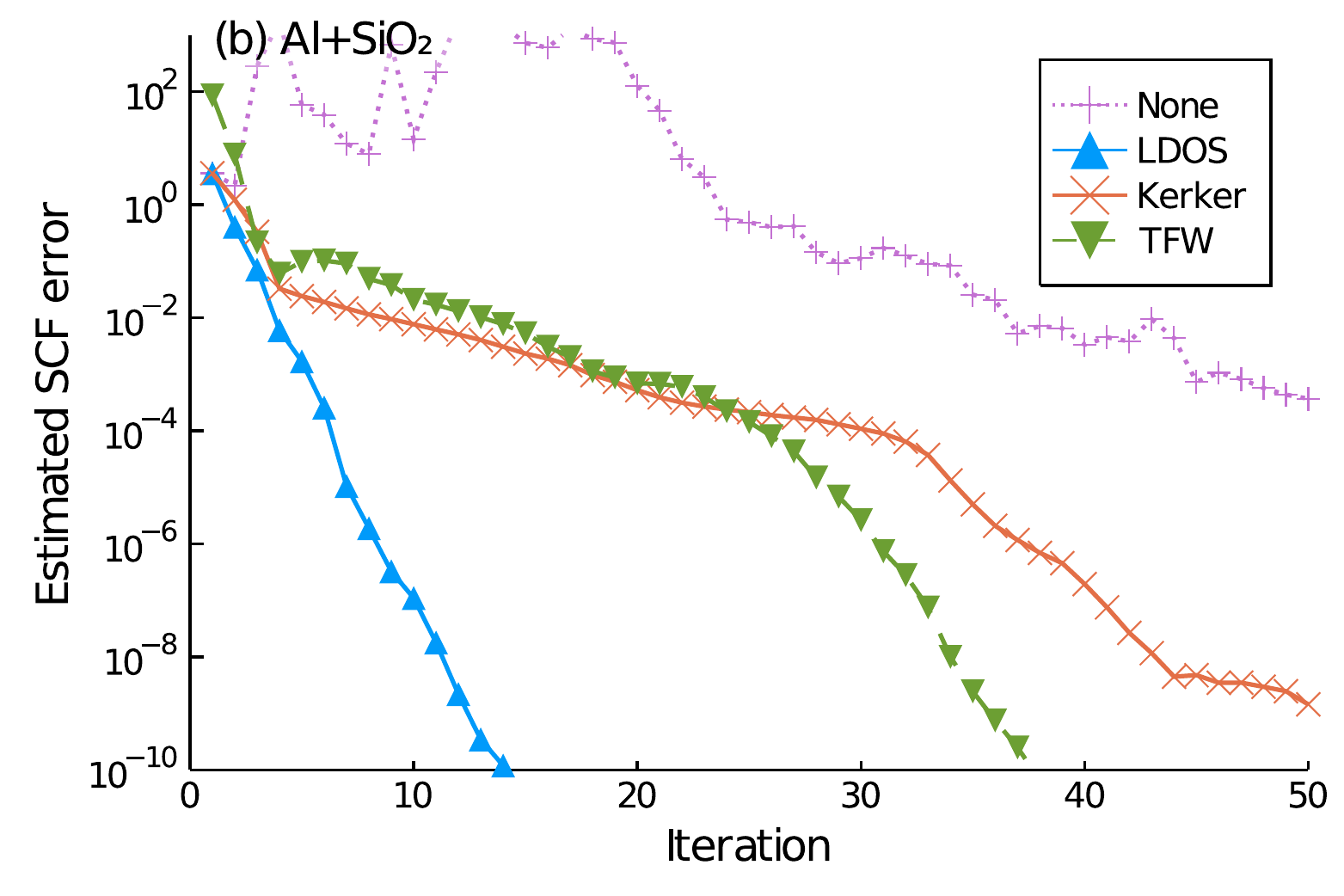}
    \caption{Convergence plots for (a) the \ce{Al}+vacuum
    and (b) the \ce{Al}+\ce{SiO2} systems with 20 repeats.}
    \label{fig:convtf}
  \end{figure}

\begin{figure}
    \centering
    \includegraphics[width=0.9\columnwidth]{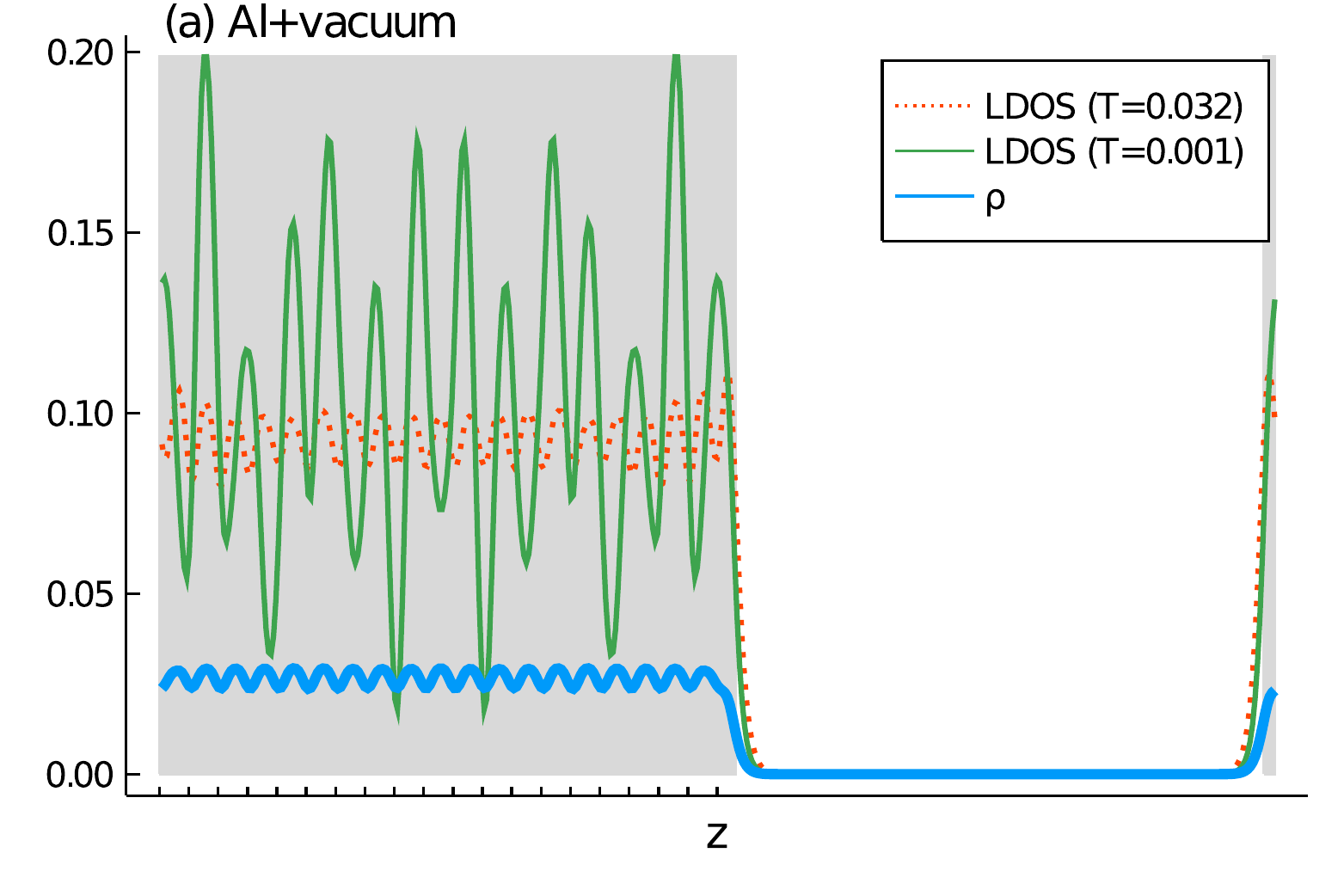}
    \includegraphics[width=0.9\columnwidth]{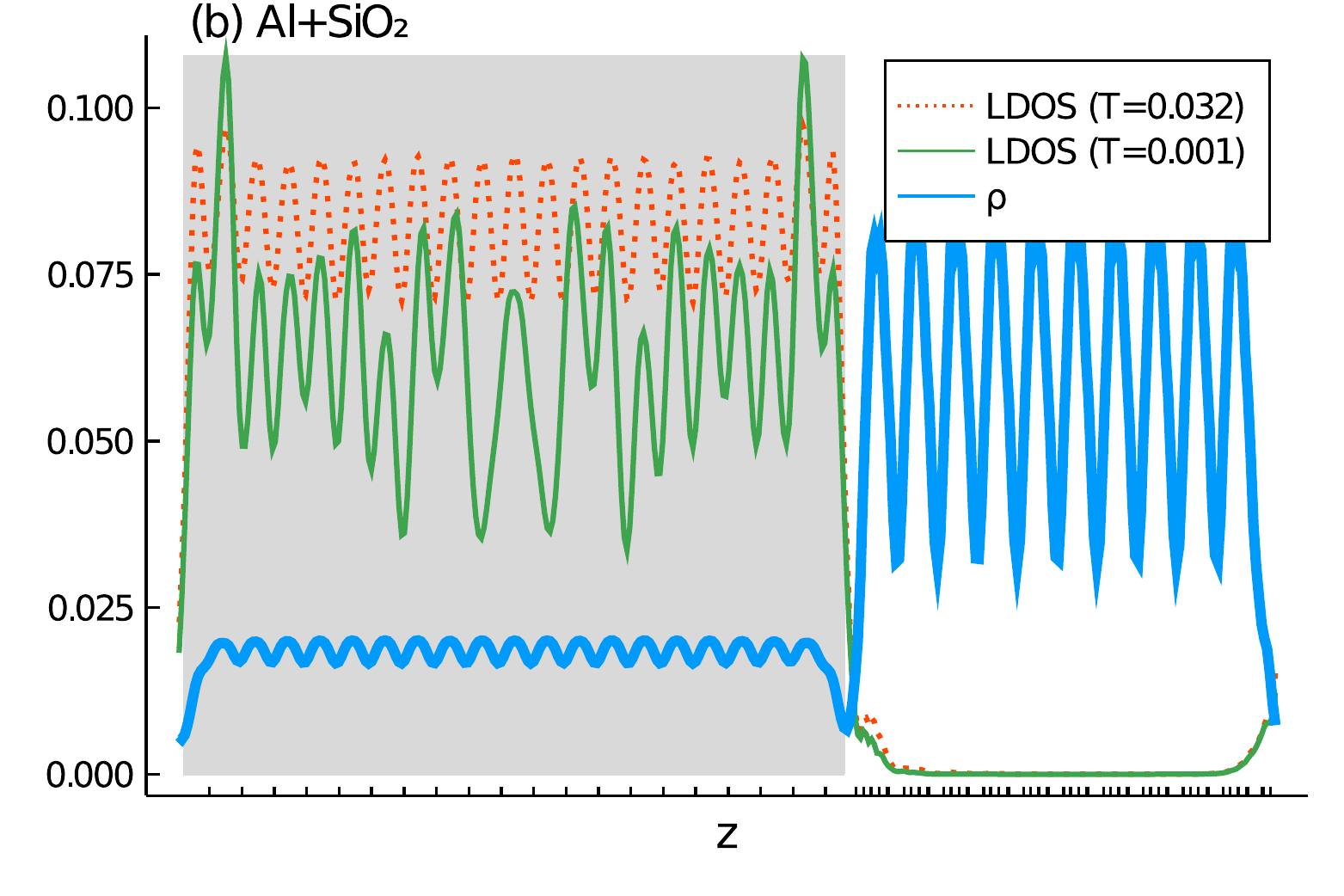}
    \caption{
        Local density of states~(LDOS) and pseudodensity $\rho$ (thick line)
        of the (a) \ce{Al}+vacuum system
        and the (b) \ce{Al}+\ce{SiO2} system.
        Values are averaged in the $xy$-plane.
        The LDOS is shown both for the temperature
        used for preconditioning and during the SCF
        ($T = 0.001$ Hartree)
        and at a higher temperature, where it is more accurate due to
        insufficient Brillouin zone sampling.
        The part of the cell occupied by aluminium
        is shown in grey background and atomic positions
        projected onto the $z$-axis are shown with a tick.
    }
    \label{fig:ldos}
  \end{figure}

In Figure~\ref{fig:convtf}, we compare our preconditioner against the
inhomogeneous TFW preconditioner of \citet{raczkowski2001thomas} on two
specific metallic systems, for which the TFW model should be adequate
in the metallic region: the \ce{Al} surface, and the \ce{Al}+\ce{SiO2}
interface. On both these systems, our LDOS preconditioner converges in
about 10 iterations. By contrast, the TFW preconditioner is less efficient, the
effect being especially pronounced in the \ce{Al}+\ce{SiO2} system
where the TFW model is very inadequate in the \ce{SiO2} region, being based
on the density instead of the LDOS for determining dielectric
properties. We illustrate this point by plotting in
Figure~\ref{fig:ldos} the pseudodensity and the LDOS in these two
systems. As is apparent, the density is drastically different in the
vacuum and the \ce{SiO2} regions, although both have very similar
screening properties, having no free electrons. By contrast, the LDOS
is able to correctly recognize the lack of free electrons in the
insulator region.

In Figure~\ref{fig:ldos}, we also plot the LDOS computed at two
different temperatures: that used in the SCF, and a higher
temperature. The LDOS at the SCF temperature is very noisy because of
the insufficient Brillouin zone sampling relative to the small
temperature used. In contrast the LDOS at the elevated temperature
is more regular, closer to the actual LDOS of this system.
Nevertheless, we found the LDOS preconditioning to be relatively insensitive
to the noise at the SCF temperature in our test cases.
Should this manifest as an issue in other calculations,
a higher smearing temperature could be used for determining the LDOS,
or the LDOS could be artificially smoothed.

  Apart from the rather artificial (but illustrative) test systems
  considered here, we have also applied the LDOS preconditioner to
  further tests taken from the \texttt{scf-xn} test
  suite~\cite{woods2019computing}. These similarly demonstrate that
  this preconditioner, with its ability to switch on and off according
  to the character of the system (metallic or not), is especially
  applicable to mixed systems containing metals, insulators and
  vacuum. For example, using a kinetic energy cutoff of
  \unit[35]{Hartree} and otherwise the computational parameters from
  \citet{woods2019computing}, we found that the LDOS
  preconditioner, gives a rapid and steady convergence to an energy
  error of $10^{-8}$ in around $10$ iterations for a variety of test
  cases such as a caffeine molecule, a slab of gold (\ce{Au40}
  supercell), or a cluster of $51$ water molecules, and is therefore
  more appropriate as a ``black-box'' preconditioner than the
  homogeneous schemes. It was however as ineffective as
  homogeneous schemes in ``complicated'' inhomogeneous metals with
  localized orbitals close to the Fermi level.

\section{Conclusion}
A novel preconditioning scheme for the self-consistent
Kohn-Sham equations based on the local density of states~(LDOS) has
been constructed and its performance evaluated in representative test
systems. We showed our scheme to be broadly applicable for large inhomogeneous systems involving metals, vacuum and
insulators.
The key idea of the proposed preconditioning scheme is to
approximate the independent-particle susceptibility $\chi_0$ by its
large-scale component \eqref{eqn:chi0ldos}, which only involves the
readily computable LDOS. The result is a preconditioner which
is simple to implement, inexpensive to compute and free of any
parameters. By nature of the LDOS the scheme automatically adapts to
the local environment in mixed systems, effectively interpolating
between a Kerker-like treatment in metallic regions and no
preconditioning where the local density of states is zero.

We believe our proposed LDOS-based preconditioning scheme to be an
important step towards a generally applicable, parameter-free and
black-box preconditioner for SCF iterations. In its present state
there are, however, a number of limitations. Firstly, as we have
discussed the LDOS is not able to distinguish insulators and
semiconductors. As a remedy we have investigated an empirical
correction by adding a dielectric model parametrized in the
macroscopic dielectric constant. The resulting preconditioner performs
very well in cases with semiconducting regions, but an appropriate
dielectric constant has to be selected.

In order to go beyond the current approach, one could try to improve
on the diagonal approximation for $\chi_0$ to represent polarizability
effects. Unlike the LDOS, the polarizability is however more expensive
to compute and cannot be easily localized \cite{sipe1978limitations},
making it difficult to develop a fully transferable preconditioning
scheme. Another limitation of our current preconditioner is that it
only focuses on preventing long-range charge-sloshing and none of the
other sources of ill-conditioning in SCF schemes (electronic phase
transitions and localized states close to the Fermi level). We intend
to tackle this in future work. We believe the idea of
approximating $\chi_{0}$ instead of directly tackling
the inverse dielectric operator to be fruitful for this endeavor as well,
because $\chi_{0}$ is susceptible to additive approximations (unlike $\varepsilon^{-1}$).
Therefore different physics (charge transfer, polarizability...)
can be summed up, resulting in a composable dielectric model.
Finally, we mention that
inexpensive approximations of the dielectric operator can be useful in
other contexts as well, such as for preconditioning geometry
optimization, in response computations, or for constructing screened
Coulomb interactions.

\section*{Acknowledgments}
This project has received funding from the
ISCD (Sorbonne Universit\'e) and from the European Research Council (ERC) under
the European Union's Horizon 2020 research and innovation program (grant
agreement No 810367).
We gratefully acknowledge stimulation discussions with E.~Cancès, X.~Gonze,
P.~Hasnip, L.~Lin, and C.~Yang.

\bibliography{literature}
\end{document}